\documentclass[twocolumn]{aastex63}

\newcommand{\nw}{nW m$^{-2}$ sr$^{-1}$}

\newcommand{\LambdaCDM}{$\Lambda$CDM}

\usepackage{amsmath}
\usepackage{natbib}
\usepackage{comment}

\submitjournal{ApJ}

\shorttitle{Exploring the PBH-\LambdaCDM~Universe}
\shortauthors{Cappelluti, Hasinger \& Natarajan}
\graphicspath{{./}{figures/}}
\begin{document}

\title{Exploring the high-redshift PBH-\LambdaCDM~Universe: early black hole seeding, the first stars and cosmic radiation backgrounds.}

\correspondingauthor{Nico Cappelluti}
\email{ncappelluti@miami.edu,guenther.hasinger@esa.int, priyamvada.natarajan@yale.edu}

\author[0000-0002-0786-7307]{Nico Cappelluti}
\affiliation{Department of Physics, University of Miami, Coral Gables, FL 33124, USA}
\affiliation{INAF – Osservatorio di Astrofisica e Scienza dello Spazio di Bologna, Via Gobetti 93/3, 40129 Bologna, Italy}

\author[0000-0002-0797-0646]{G\"unther Hasinger}
\affiliation{ European Space Astronomy Centre (ESA/ESAC)
E-28691 Villanueva de la Ca\~nada, Madrid, Spain
}
\author[0000-0002-5554-8896]{Priyamvada Natarajan}
\affiliation{Department of Astronomy, Yale University, 52 Hillhouse Avenue, New Haven, CT 06520, USA}
\affiliation{Department of Physics, Yale University, P.O. Box 208121, New Haven, CT 06520, USA} 
\affiliation{Black Hole Initiative, Harvard University, 20 Garden Street, Cambridge MA 02138, USA}

\
%% Note that the \and command from previous versions of AASTeX is now
%% depreciated in this version as it is no longer necessary. AASTeX 
%% automatically takes care of all commas and "and"s between authors names.

%% AASTeX 6.3 has the new \collaboration and \nocollaboration commands to
%% provide the collaboration status of a group of authors. These commands 
%% can be used either before or after the list of corresponding authors. The
%% argument for \collaboration is the collaboration identifier. Authors are
%% encouraged to surround collaboration identifiers with ()s. The 
%% \nocollaboration command takes no argument and exists to indicate that
%% the nearby authors are not part of surrounding collaborations.

%% Mark off the abstract in the ``abstract'' environment. 
\begin{abstract}
We explore the observational implications of a model in which primordial black holes (PBHs) with a broad birth mass function ranging in mass from a fraction of a solar mass to $\sim$10$^6$ M$_{\odot}$, consistent with current observational limits, constitute the dark matter component in the Universe. The formation and evolution of dark matter and baryonic matter in this PBH-\LambdaCDM~Universe are presented. In this picture, PBH DM mini-halos collapse earlier than in standard \LambdaCDM, baryons cool to form stars at $z\sim15-20$, and growing PBHs at these early epochs start to accrete through Bondi capture. The volume emissivity of these sources peaks at $z\sim20$ and rapidly fades at lower redshifts. As a consequence, PBH DM could also provide a channel to make early black hole seeds and naturally account for the origin of an underlying dark matter halo - host galaxy and central black hole connection that manifests as the $M_{\rm bh}-\sigma$ correlation. To estimate the luminosity function and contribution to integrated emission power spectrum from these high-redshift PBH DM halos, we develop a Halo Occupation Distribution (HOD) model. In addition to tracing the star formation and reionizaton history, it permits us to evaluate the Cosmic Infrared and X-ray Backgrounds (CIB and CXB). We find that accretion onto PBHs/AGN successfully accounts for the detected backgrounds and their cross-correlation, with the inclusion of an additional IR stellar emission component. Detection of the deep IR source count distribution by the JWST could reveal the existence of this population of high-redshift star-forming and accreting PBH DM.

\end{abstract}

%% Keywords should appear after the \end{abstract} command. 
%% See the online documentation for the full list of available subject
%% keywords and the rules for their use.
\keywords{editorials, notices --- 
miscellaneous --- catalogs --- surveys}

%% From the front matter, we move on to the body of the paper.
%% Sections are demarcated by \section and \subsection, respectively.
%% Observe the use of the LaTeX \label
%% command after the \subsection to give a symbolic KEY to the
%% subsection for cross-referencing in a \ref command.
%% You can use LaTeX's \ref and \label commands to keep track of
%% cross-references to sections, equations, tables, and figures.
%% That way, if you change the order of any elements, LaTeX will
%% automatically renumber them.
%%
%% We recommend that authors also use the natbib \citep
%% and \citet commands to identify citations.  The citations are
%% tied to the reference list via symbolic KEYs. The KEY corresponds
%% to the KEY in the \bibitem in the reference list below. 

\section{Introduction} \label{sec:intro}

Dark matter (DM) represents the most abundant form of matter in the Universe and dominates the dynamics of collapsed objects. It also offers the scaffolding within which all visible matter is structured into galaxies. Thus far, in the context of the cold dark matter paradigm, it has been widely assumed that DM exists in the form of still unknown particles that interact primarily through gravity and perhaps through weak interactions \citep[e.g.][]{2010ARA&A..48..495F}. However, despite several decades of targeted experimental searches aimed at uncovering weakly interacting massive particles as potential dark matter candidates, these efforts have all come up empty. At present there seem to be no sign of particle DM candidates in the mass interaction cross-section parameter space and energy ranges where they have been predicted \citep{Baudis2012,Boveia+2018}. 

Meanwhile, the discovery of gravitational waves from merging black hole (BH) binaries by LIGO and VIRGO reveal surprisingly large masses for the individual merging BHs with, on average, low pre-merger spins. Typical inferred masses of the merging sources are higher than expected from astrophysical formation channels \citep{2016PhRvX...6d1015A}. Consequently, the community revived the hypothesis originally proposed by \citet{1971MNRAS.152...75H} that DM could be constituted by Primordial Black Holes (PBHs) that formed in the infant Universe \citep{karsten,Carr2003,2016PhRvL.116t1301B,kash16,2017PDU....15..142C,karsten2}. 

Early models assumed that all dark matter is comprised of PBHs that formed with a monochromatic mass function, but observational constraints firmly rule out this hypothesis \citep[e.g.][]{2019EPJC...79..246B}. Later work and further refinements, notably by \citet{Carr+2019}, \citet{belindo}, \citet{Carr} and \citet{Carr-Bellido2021} showed that DM PBHs can, in principle, have a broad birth mass spectrum ranging from $10^{-10}\,-\,10^{7}\,M_{\odot}$; while accounting for all the DM without violating current observational constraints. In their model, PBHs are created in the early Universe during QCD phase transitions (around 100 MeV that corresponds to $\sim 10^{10}\,K$) involving different particle families freezing out of the primordial quark-gluon plasma within the first two seconds after the inflationary phase. When $W^{+/-}$, Z bosons, baryons, pions are created, and e$^+$e$^-$ pairs annihilate, they leave an imprint in form of a significant reduction of the sound speed at the corresponding phase transitions, thereby causing regions of high curvature to collapse and form PBHs. The typical mass scale of these PBHs is defined by the size of the horizon at the time of the corresponding phase transition. In this model, four distinct populations of PBHs in a wide mass range are expected to form: planetary mass black holes at the $W^{+/-}$-Z transition; PBHs of around the Chandrasekhar mass when the baryons (protons and neutrons) form from 3 quarks; PBHs with masses of order 30 M$_{\odot}$ (these correspond to the suggested LIGO black holes), when pions form from two quarks; and finally PBHs with masses corresponding to those of supermassive black holes (SMBHs) with $M \geq 10^6\,M_{\odot}$ that form at the e$^+$e$^-$ annihilation. If PBHs form with a broad mass distribution, the DM they constitute is expected to strongly cluster, which would help alleviate some of the more stringent observational constraints on the allowed contribution of PBHs to the dark matter \citep{2017PDU....15..142C,2019EPJC...79..246B} budget. 

Intriguingly, an excess of small-scale DM substructure compared to CDM predictions has been recently reported from gravitational cluster lensing studies with the deepest HST observations  \citep{2020Sci...369.1347M}. Clustering of DM in excess of what is predicted by the standard WIMP CDM paradigm as expected with PBH DM, could possibly account for this discrepancy. This excess concentration of mass on small scales is revealed in the discrepancy between the observed and predicted event rates for Galaxy-Galaxy Strong Lensing (GGSL) events. The internal structure of subhalos with masses $\sim 10^{11}\,M_{\odot}$ are implicated for these GGSL events, and \LambdaCDM~ simulations simply do not produce enough subhalos in this range with the requisite central concentrations. 

In addition to possibly providing a simple resolution of the nagging DM problem, PBHs it appears could also serve to account for early massive black hole seed formation and  address the intriguing origin of the SMBHs with mass of the order ${10}^{10}$ M$_{\odot}$ powering detected luminous quasars already in place by $z\,>\,7$ when the Universe was $\,<\,$0.8 Gyr old \citep[see e.g.][]{lodatoPN2006,2007ApJ...665..187L}. 

Quite a number of other recent observational results also strengthen the conjecture that PBHs could contribute to the overall DM budget. The latest GWTC-2 catalogue of LIGO-Virgo-KAGRA gravitational merger events \citep{2021PhRvX..11b1053A} has widened the observed BH mass distribution considerably. In particular, it includes the most massive merger detected as yet GW190521 \citep{Carr+2019,clesse2021gw190425,2020PhRvL.125j1102A}, in which at least one of the two components is more massive than the upper mass gap expected for pair instability supernovae (SN), and thus could signal a PBH origin \citep{clesse2021gw190425,2021PhRvL.126e1101D}. It also includes the event GW190814 \citep{2020ApJ...896L..44A}, in which one of the components likely falls into the lower SN mass gap between neutron stars and BH, and which has a surprisingly large mass ratio of $\sim$~1:9, not entirely easily compatible with known astrophysical production channels \citep{2021PhRvL.126e1101D}. \cite{2021PhRvD.103b3026W} analyse the whole GWTC-2 catalogue and conclude that the observed event rate is fully consistent with the assumption that all LIGO-VIRGO detected merging BHs are of primordial origin, contributing a fraction of $f_{PBH}\,\sim\,0.3$\% to overall the DM budget in the  mass range 1$<$M$_{BH}<$100 M$_{\odot}$. 

New results from the 5-year OGLE micro-lensing campaign \citep{2019PhRvD..99h3503N} present the discovery of a sizeable population of long-duration micro-lensing events, which, together with Gaia parallaxes, point to the presence of putative PBHs in the mass range 1--10 M$_{\odot}$ \citep{2020A&A...636A..20W}. Another mass-gap BH candidate was recently discovered in the nearby nearly edge-on ellipsoidal variable binary star V723 Mon  \citep{2021MNRAS.504.2577J}. OGLE has also detected 6 ultrashort-timescale microlensing events, which may, in fact, indicate the existence of planetary mass PBHs \citep{2019PhRvD..99h3503N}.  The NANOGrav pulsar timing observatory has recently reported interesting upper limits on the stochastic gravitational wave background in the nano-Hertz band in their data, which does not show statistically significant quadrupolar spatial correlations expected for a cosmic gravitational wave background \citep{2020ApJ...905L..34A} yet. Meanwhile, the Parkes Pulsar Timing Array (PPTA) collaboration has also reported a tentative detection consistent with the NANOGrav finding \cite{PPTA2021}. However, the derived upper limit they provide is compatible with several PBH formation processes operating over a wide mass range \citep{2021PhRvL.126d1303D,2021PhRvL.126e1303V,2021PhLB..81336040K,2020arXiv201003976D,2021PhLB..81436097S}. Recent indications from GRB microlensing constraining f$_{PBH}\sim 3\times10^{-3}$ for $10^6$ M$_\odot$ PBHs \citep{2021arXiv210500585K}, are consistent with the mass spectrum considered here. Finally, microlensing events from multiply lensed quasar images have been interpreted to indicate that the dark matter in galaxy haloes and clusters could be potentially composed of PBH with masses of around one solar mass \citep{2020A&A...633A.107H,2020A&A...643A..10H}.  All of these currently reported observational constraints and bounds are plotted in Fig.~1. However, we note that these are mostly derived for a monochromatic initial mass spectrum for PBHs and not for the broader mass spectrum we investigate in here this work. 

In addition to this suggestive circumstantial evidence, there are several other open problems in cosmology for which the PBH-DM hypothesis might proffer explanatory power \citep{Carr+2019}. The exciting implications for structure formation and evolution at early cosmic epochs in a PBH-DM Universe motivates our current detailed exploration of this model. A notable unsolved mystery in observational astrophysics is the origin of the large scale excess fluctuations in the unresolved Cosmic Infrared Background (CIB) discovered by \citet{2005Natur.438...45K}, confirmed by \citep{k2007ApJ...654L...5K,2012Natur.490..514C,2012ApJ...753...63K} and their coherence with the Cosmic X-ray Background (CXB) \citep{cap13,cap17,2016ApJ...832..104M,li18,m2011ApJ...742..124M}. Careful modeling of the shape of these fluctuations, suggests two possible origins: (i) that they are consistent with being produced in the young Universe from early stellar emission or (ii) from more local intra-halo light. However, the coherence with the CXB is even more intriguing because it suggests an origin from accretion powered emission from the first black holes, believed to be the progenitors of observed SMBHs. There is no clear consensus on how these SMBH seeds formed and several possible scenarios have been invoked, including the collapse of Population III stars \citep[see e.g][]{k1983MNRAS.205..955K} and Direct Collapse Black Holes \citep[DCBH, see reviews by][]{Volonteri2012Sci,Natarajan2014}. \citet{ricarte19} showed that neither DCBHs nor Pop III stars could in fact produce the required amount of radiation to explain the observed cross-power spectrum and excess. In addition, the origin of the empirical scaling between the masses of central SMBHs and the stellar velocity dispersion of their host galaxies and dark matter haloes is poorly understood at present, and whether they arise as a result of the seed formation process (nature) or emerge over time as a consequence of growth and assembly in tandem (nurture) is debated \citep{LodatoNatarajan2006}.

Another potential problem arises for models of emission from Pop III stars in the early Universe. Extrapolating the star formation density observed at $z\,<\,8$ to higher redshifts, \cite{cooray12reio} point out that the expected emissivity of Pop III stars falls short in re-producing the required amplitude of the CIB power spectrum. \cite{helgason16} studied the physical conditions required of early star forming galaxies to produce enough flux to explain the observed CIB flux and power spectrum and note that in order to reproduce the observed signal, stars need to form either with an unreasonable baryon conversion efficiency $f_{\star}>0.1$ and/or with extremely top-heavy and tilted Initial Mass Functions (IMFs) with all stars being born with masses larger than 500 M$_{\odot}$. Even considering other proposals, problems still persist in reconciling the observed source number counts and Planck reionization constraints \citep{yue+2013}.

According to \citet{kash16}, if DM is made of PBH then we should expect that at redshifts $z\,>\,15$  they could accrete a substantial amount of matter to emit enough in the IR (rest frame UV) and X-ray wavelengths to significantly contribute to the CIB and CXB. \citet{afshordi} pointed out that in a PBH-DM \LambdaCDM~ cosmology, the fraction of collapsed halos at high redshifts, specifically at $z\,>\,7$, is significantly higher than in the standard \LambdaCDM~ Universe. Due to this excess, even a modest efficiency in converting baryons into radiation could give rise to significantly higher emissivity either through accretion or star formation. 

\begin{figure*}[!htp]
    \centering
    \includegraphics[width=0.75\textwidth]{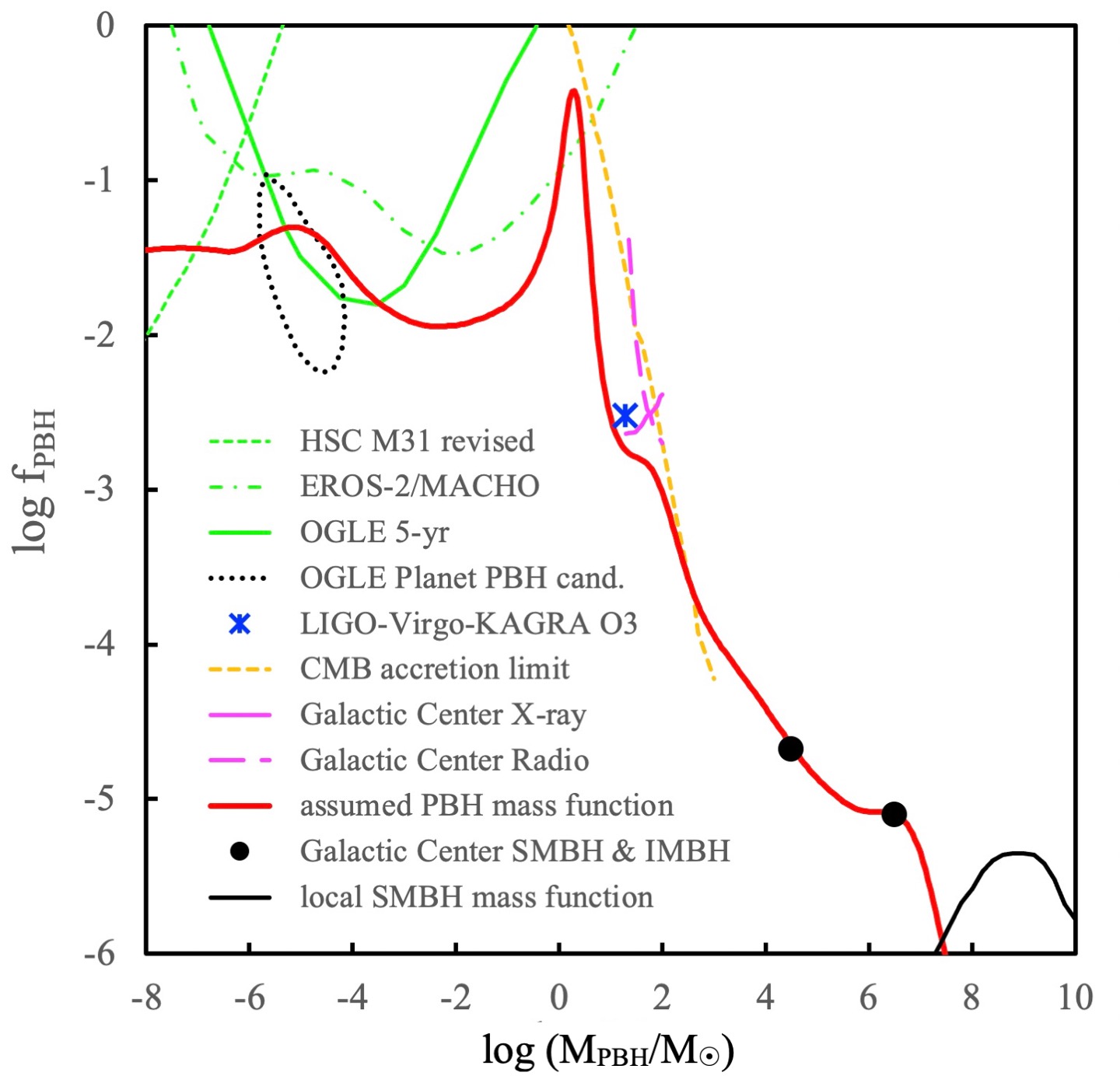}
    \caption{The PBH mass spectrum (thick red line) assumed in this work \citep{has}, along with a number of over-plotted current observational constraints  mostly derived for a monochromatic initial mass function: (i) microlensing limits from the Subaru M31 survey \citep{2019NatAs...3..524N} updated by \citet{2020PhRvL.125r1304K}; (ii) from the EROS-2/MACHO survey \citep{2007A&A...469..387T} and (iii) from the OGLE 5-year survey \citep{2019PhRvD..99h3503N} are shown in green dashed, solid and dot-dashed lines, respectively. The dotted black line shows the 95\% confidence region, assuming that the 6 ultrashort-timescale microlensing events in the OGLE data are due to planetary mass PBHs \citep{2019PhRvD..99h3503N}. The blue star indicates the PBH fraction derived from the assumption that all BH mergers observed in the third observing run (O3) of the LIGO-Virgo-KAGRA Collaboration are PBHs \citep{2021PhRvD.103b3026W}. The CMB accretion limits from \cite{2017PhRvD..96h3524P} are shown as the orange dashed line. Multi-wavelength limits from models of the Galactic Center \citep{2019JCAP...06..026M} are shown in magenta for X-ray (solid) and radio (dashed) observations. The two black circles correspond to 10  intermediate-mass black holes (so far 5 have been observed) and the SMBH in the Galactic Center \citep{has}. Finally, the local SMBH mass function derived from inactive SMBHs hosted in galactic nuclei \citep{natarajan+2009,2013CQGra..30x4001S} is shown as the black curve at 10$^{7-10}$ M$_{\odot}$.}
    \label{fig:fPBH}
\end{figure*}

Therefore, computing the detailed history of emission from star formation formation and black hole accretion in a PBH-DM Universe over cosmic time is warranted. In a recent paper, \citet{has} predicted the amount of extra-galactic background flux produced by accretion onto PBHs by Bondi capture via advection dominated flows (ADAF). In this scenario, PBHs account for all the DM with a broad mass spectrum peaking near the Chandrasekar mass of 1.4 M$_{\odot}$, following \citet{belindo} (see figure \ref{fig:fPBH}). In this event, PBH accretion can produce about 1\% of the total [0.5-10] keV X-ray background and about 0.5\% of the CIB necessary to explain the observed large scale, near infrared surface brightness fluctuation power spectrum excess. This emission arises from the recombination process that is spread over redshift from $z\,\sim\,1100$, peaking at $z\,\sim\,20$ and rapidly fading by $z\,\sim\,7-8$. This amount of radiation produced by PBHs and its redshift distribution appears to be consistent with the overall shape of the CIB power spectrum and the intensity required to explain the observed large scale cross-correlation excess of the CIB and CXB from known  populations. These very same accreting PBHs, it turns out, can also produce sufficient radio background emission at high redshift to explain the surprisingly strong sky-averaged red-shifted 21-cm line signal observed by {\em EDGES} \citep{has,2018Natur.555...67B}. The signal is reported to be centered at a frequency around 78 MHz and covers a broad range in redshift $z\,=\,15-20$ \citep{2018Natur.555...67B}. 

Here, we investigate in detail the contribution of PBHs and consequences thereof at cosmic dawn for the mass assembly history of dark matter; star formation and re-ionizaton history; BH growth and the production of cosmic radiation backgrounds. While, the CXB vs. CIB cross-correlation excess might not be produced at high redshift \citep[see e.g.][]{2012Natur.490..514C} here we model and investigate in addition to the global properties enumerated above, the observed auto- and cross-power spectra of the CIB and CXB using the predictions of \citet{has}. We adopt the standard Press-Schechter formalism to derive the DM halo mass function and combine this with a Halo Occupation Distribution (HOD) approach to compute the clustering strength of PBHs in DM halos. In this paper, we assume a PBH-DM \LambdaCDM~(hereafter referred to as the PBH-\LambdaCDM~model) cosmology with parameters $\Omega_{\Lambda}$= 0.7, $\Omega_{m}$= 0.3 and H$_{0}$=67h$^{-1}$ km/s/Mpc when needed. 

\begin{figure*}[t]
    \centering
    \includegraphics[width=0.7\textwidth]{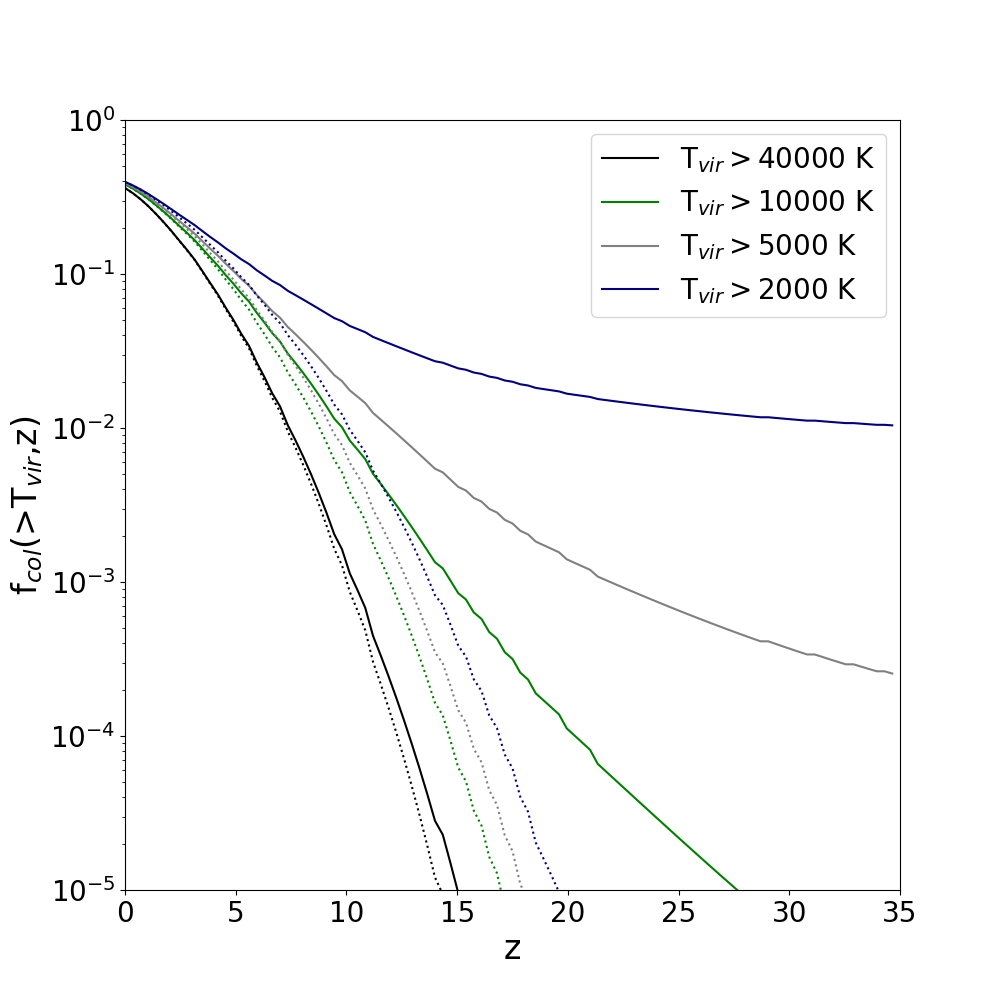}
    \caption{Comparison of the fraction of collapsed halos with masses greater than $M$ as a function of redshift in the PBH-\LambdaCDM~ and \LambdaCDM~ cosmologies, shown as continuous and dotted lines, respectively. Black, green, grey and navy lines represent halos with $T_{vir}\,>\,40000$ K; $T_{vir}\,>\,10000$ K; $T_{vir}\,>\,5000$ K and $T_{vir}\,>\,2000$ K (corresponding at $z=10$ for instance, to halo masses M$_h$=2.6$\times$10$^6$ M$_\odot$, M$_h$=1.0$\times$10$^7$ M$_\odot$, M$_h$=2.9$\times$10$^7$ M$_\odot$ and M$_h$=2.4$\times$10$^8$ M$_\odot$), respectively. The excess formation rate for low temperature (mass) halos with virial temperatures in the 2000-10000 K range in the PBH-\LambdaCDM~ cosmology is clearly evident and pronounced at all $z\,>\,10$. This is a key feature that distinguishes the PBH-\LambdaCDM~ and \LambdaCDM~ cosmological models.}
    \label{fig:halo_mass}
\end{figure*}

\section{Modeling Structure Formation in the PBH-\LambdaCDM~ Universe}

\subsection{Mass Power Spectrum and Mass Function of Collapsed Halos}
 
 The linear power spectrum of density fluctuations in the PBH Cold Dark Matter Universe has been described by \citet{afshordi} as follows: 

\begin{equation}
P_{lin}(k)=P_{\Lambda CDM}(k)+P_{Poiss}(k),
\label{eq:PBH}
\end{equation}
where $k$ is the spatial frequency,  P$_{\Lambda CDM}$ is the power spectrum of Dark Matter and P$_{Poiss}$ is an additional component introduced by the discrete nature of PBH dark matter \citep{M1975A&A....38....5M}. This component has little or no effect on large scales but becomes dominant on very small scales in the form of shot noise. Note that with this assumed form, the PBH component is not affected by bias. The Poisson piece can be approximated as:

\begin{equation}
P_{Poiss}=\frac{9}{4}(1+z_{eq})^2n_{BH}[g(z)]^{-2}
\end{equation}

where $g(z)$ is the linear growth factor of fluctuations from $z$ to today, with $g(0) = 1$; and where n$_{BH}$ = $\frac{f_{BH}}{M_{BH}}$ is the number density of PBHs. As PBHs have a wide range of birth masses, the Poissonian component needs to be evaluated as the density weighted number density of PBHs, which in our assumed case, can be safely represented by the mass scale M$_{BH}$ = 1.4 M$_{\odot}$ which corresponds roughly to the peak observed in the mass spectrum plotted in Fig.~1. Per standard methodology, with P$_{M}(k,z)$ the matter density fluctuation power spectrum written out as function of the spatial frequency $k$ and redshift, we can evaluate the rms density contrast over spherical region of co-moving radius r$_M$ and of mass M(r$_M$) as: 
$$\sigma_M(z)\,=\,\left[\int P_{M}(k,z) W_{TH}(kr_M)k^2dk\right]$$ 
where W$_{TH}$ is the standard top-hat window function.  Then, using the Press-Schechter formalism \citep[PS,][]{ps}, we evaluate the fraction of mass in a given volume that is contained in halos of a given mass at redshift $z$, which is given by: 
 \begin{equation}
 f_{halo}(>M_h,z)=\frac{1}{2}\,erfc\left(\frac{\delta_{col}}{\sqrt{2}\sigma_{M}(z)}\right),
 \end{equation}
 where $\delta_{col}$ is the over-density required for spherical top-hat collapse. 
 
 In Fig.~\ref{fig:halo_mass}, we compare the fraction of collapsed halos as function of $z$ in PBH-\LambdaCDM~ and in the classic \LambdaCDM~ cosmology, using the temperature as the proxy for mass. Conversion from one to the other is straight forward, for example, at $z\,=\,10$, the halo masses corresponding to the temperature range studied span from $M_h\,=\,2.6\,\times\,10^6$ M$_\odot$ to $M_h\,=\,2.4\,\times\,10^8$ M$_\odot$. Notably, we see that in the early Universe, at $z\,>\,15$, the PBH-\LambdaCDM~ model predicts a significantly larger number of mini-halos (i.e. $M_h\,\sim\,10^6\,M_{\odot}\,-\,10^7\,M_{\odot}$) that correspond to $T_{\rm vir}\,>\,2000$ K, than \LambdaCDM. In the PBH-\LambdaCDM~ scenario, mini-halos wherein the first stars form through molecular cooling assemble much earlier than in the \LambdaCDM~ Universe, shifting the beginning of star formation to a much earlier epoch. Simultaneously, by forming earlier, the first light emitted by these highly biased halos imprints a stronger clustering signal in the power spectrum of the diffuse background fluctuations. This early activity would also inject ionizing photons into the IGM earlier than in the standard star formation models assumed for \LambdaCDM. We compute the detailed consequences of astrophysical processes that occur in this excess population of early mini-halos in the following sections.

\section{Physical Processes in the PBH-\LambdaCDM~ Universe}

\subsection{Early Star Formation}

First, we evaluate how the significantly higher mini-halo space density at high-z predicted in the PBH-\LambdaCDM~ scenario noted above impacts the star formation rate. To do so, we follow the prescription of \citet{gb06}, as done previously by \citet{helgason16}. The virial temperature of a halo is related to its mass via:
\begin{eqnarray}
M=10^8 M_{\odot}(\frac{\mu}{0.6})^{-3/2}(\frac{T_{vir}}{10^4 K})^{3/2}   (\frac{1+z}{10})^{-3/2}
\end{eqnarray}
where $\mu$ is the mean molecular weight. Here we adopt $\mu$=1.2 consistent with the fact that the Universe was mostly neutral at high-$z$. According to this prescription, the star formation in early galaxies is taken to be proportional to the halo collapse rate at a given redshift as:
\begin{equation}
    \dot{\rho_{\star}}(M_h,z)=f_\star\frac{\Omega_b}{\Omega_M}\frac{d}{dt}M\frac{dn}{dM_h}(>M_{min},t).
    \label{eq:sfr}
\end{equation}
where f$_{\star}$ is baryon conversion efficiency (i.e. the fraction of baryons that are converted into stars) and M$_{min}$ is the minimum halo mass that is permitted to shock heat to T$_{vir}$ and consequently allow gas cooling. Consistent with the treatment of \citet{helgason16}, we consider two ranges of halo temperatures, each with its own dominant coolant. One population, with $10^3\,K\,<\,T_{vir}\,<\,4\,\times\,10^4$ K where most of the cooling is due to molecular Hydrogen and the other atomic cooling halos with $T_{vir}\,>\,4\,\times\,10^4$ where cooling is mostly due to mono-atomic Hydrogen. While the efficiency of star formation, f$_\star$ at high-z is unknown (it is scarcely observationally constrained even at more recent epochs), guided by the work of \citet{sun} and \citet{miro}, we model the relation between f$_\star$ and M$_h$ as: 
\begin{equation}
    f_\star(M_h)=\frac{0.05}
    {\large(\frac{M_h}{2.8\times10^{11}}\large)^{0.49}+  {\large(\frac{M_h}{2.8\times10^{11}}\large)^{-0.61} 
    }}+f_{\star_{min}},
\end{equation}
where f$_{\star_{min}}$ is floor for star formation efficiency at low mass. Unlike \cite{helgason16}, who assumed an universal value for f$_\star$, here we adopt a M$_h$ dependent value with f$_{\star_{min}}$, that is left as a free parameter. 

While in a PBH-\LambdaCDM~ Universe mini halos form early, star formation does not necessarily occur in them promptly but is in fact initially suppressed. According to criteria for star formation studied by \citet{stacy, tse,fialkov} star formation occurs only when baryons can collapse and fragment within the halo. At very high-z, this might not be possible if the circular velocity of the halo $v_{\rm cool}$ \citep[computed using the analytic prediction of][]{fialkov} is smaller than the relative streaming velocity v$_s$ of the IGM and dark matter. If $v_{\rm cool}(M_h,z)$ is the circular velocity (or cooling velocity) of a halo of mass M$_h$ at redshift $z$, we define a function $\Theta(M_h,z)$ which is 0 when $v_{\rm cool}(M_h,z)\,<\,v_s$ and 1 of $v_{\rm cool}(M_h,z)\,>\,v_s$. Basically, this step function is a simplified representation of the trigger probability of star formation in a halo. Note that $v_{\rm cool}(M_h,z)$ depends on the IGM temperature and therefore also on the ionization history. As this is also unknown at present, for the purpose of our calculation in this paper, we adopt the fiducial best-fit values proposed by \citet{fialkov}.  It is worth noting that \citet{k2021PhRvL.126a1101K} studied the timing of halo collapse in a PBH-\LambdaCDM~ Universe in the case of a monochromatic PBH mass function. While this proposal would be 
more pertinent for the physics of PBH-DM, with our model we will adapt the simplified approach described in this section because of the large number of concurring variables that might provide a different star formation history (see below). 

With this assumption, the star formation density becomes: 
\begin{equation}
    \dot{\rho_{\star}}(M_h,z)=f_\star(M_h)\frac{\Omega_b}{\Omega_M}\frac{d}{dt}M\frac{dn}{dM_h}(>M_{min},t)\Theta(M_h,z).
    \label{eq:sfr2}
\end{equation} 
As the very first star forming halos in the Universe are pristine and unpolluted by metals, it is expected that the first stars to form in these Pop III mini-halos might be substantially more massive, hotter and bluer than the subsequent generation of stars. Pop III stars are therefore expected to be the main constituents of these lower mass mini-halos until, after a few million years, they explode producing supernovae that pollute the environment with metals thereby modifying the initial mass function for the next generation of star formation. To implement this scenario, we include a function motivated by cosmological simulations of the early star formation reported in \cite{gb06}, that takes into account the metallicity evolution of halos with a parameter, $p_{\rm pris}(z)$, that represents the fraction of pristine, metal-free halos as function of redshift. We use the model of \citet{gb06} that demarcates the regimes dominated by Pop II and Pop III halos, and where the star formation density is taken to be:
\begin{equation}
\begin{split}
\dot{\rho_\star}_{Pop III}(M_h,z)=\dot{\rho_{\star}}p_{pris}(z) \\
\dot{\rho_\star}_{Pop II}(M_h,z)=\dot{\rho_{\star}}\left[1-p_{pris}(z)\right].
\end{split}
\end{equation}

As the true metallicity evolution of these halos is also unknown, in order to explore the parameter space exhaustively, here we assume three different scenarios by imposing a redshift cut at which a specified fraction of each family of halos is 50\% enriched: $z_{1/2}\,\sim\,=\,$15, 13 and 11 motivated by the excess number of PBH-DM halos that are available compared to \LambdaCDM~ at these epochs, as shown in Fig.~ 2. Going forward, we refer to these three options as {\bf Early}, {\bf Mid} and {\bf Late} enrichment scenarios, respectively. 

\subsection{Flux production rate from accretion onto PBH-DM}

After setting up the framework for including star formation, we now proceed to compute the emission from accretion onto PBH-DM. In order to evaluate the flux production rate of PBHs, we use the \citet{has} PBH-DM model to evaluate the luminosity of each halo of mass M$_h$ and apply the appropriate bolometric and $k$-corrections to estimate it in all photo-metric bands of interest. We note that these factors are typically of the order of $\sim\,10$\% and are derived, at each redshift from an accretion model - in this case, the ADAF model that aptly describes the expected obscured accretion mode at these early epochs. If $f_{\rm DM}(M_{\rm BH})$ is the fraction of DM in PBHs of a given mass, the  occupation fraction of a PBH of mass $M_{\rm BH}$ in a halo of mass M$_h$ is:
\begin{eqnarray}
n_{\rm BH}(M_h,M_{\rm BH}) = f_{\rm DM}(M_{\rm BH})\frac{M_{h}}{M_{\rm BH}}
\label{eq:npbh}
\end{eqnarray}

 The luminosity of each PBH of mass M$_{\rm BH}$ is then $L_{\rm BH}=\dot{m}\eta L_{\rm Edd},$ where $\eta$ the radiative efficiency is $\sim$ 0.1. Following \citet{has} $Log(\dot{m})$ is a function of the BH mass and redshift, peaking at around $Log(\dot{m})\,\sim\,-1.6$ at $z\,>\,20$ where the relative streaming velocity between DM and the IGM reaches its minimum value, as demonstrated in high-resolution cosmological simulations by \cite{Stacy+2011}. We refer the reader to \citet{has} and \cite{Stacy+2011} for elaboration on details of the emission properties of PBHs and coupling between baryons and DM at the earliest epochs respectively.

Once the accretion luminosity has been evaluated, the halo luminosity from PBHs can be computed and is given by:
\begin{equation}
L_{\nu_{halo}}(M_h,z)=\int f_{DM}(M_{BH})\,L_{\nu_{BH}}(M_{BH},z)\,M_h\,dM_{BH},
\end{equation}
where L$_{\nu_{BH}}$ is the normalized bolometric PBH luminosity. The halo, PBH Conditional Luminosity Function (CLF) is then given by:
\begin{eqnarray}
\phi_{L_{\nu_{BH}}}(M_h,z)=\frac{dN}{dM_{h}}(z)
\end{eqnarray}
with $\frac{dN}{dM_{h}}(z)$ being the halo mass function \citep{ShethTormen01}. The flux production rate is then simply:
\begin{eqnarray}
f_{\nu_{i}}(M_h,z)=\frac{L_{\nu_{BH}}}{4\pi D_L(z)^2}
\end{eqnarray}
where $D_L(z)$ is luminosity distance. 

The peak of the expected emission from these mini-halo populations is centered around $z\,\sim\,20$, and it extends out to $z\,\sim\,100$ in the X-ray. The bulk of this emission arises from low mass ($10^{5-6}\,M_{\odot}$) halos while no flux is observed in the IR at $z\,>\,40$ due the Lyman absorption. This early generation of the CXB in the PBH-\LambdaCDM~ model has important implications for direct collapse of gas required to form massive BH seeds in conventional \LambdaCDM .

\subsection{The formation and growth of Mini-Quasars \& AGN}

\begin{figure}[t]
\begin{center}
\hbox{\hspace{-1cm} \includegraphics[width=0.6\textwidth]{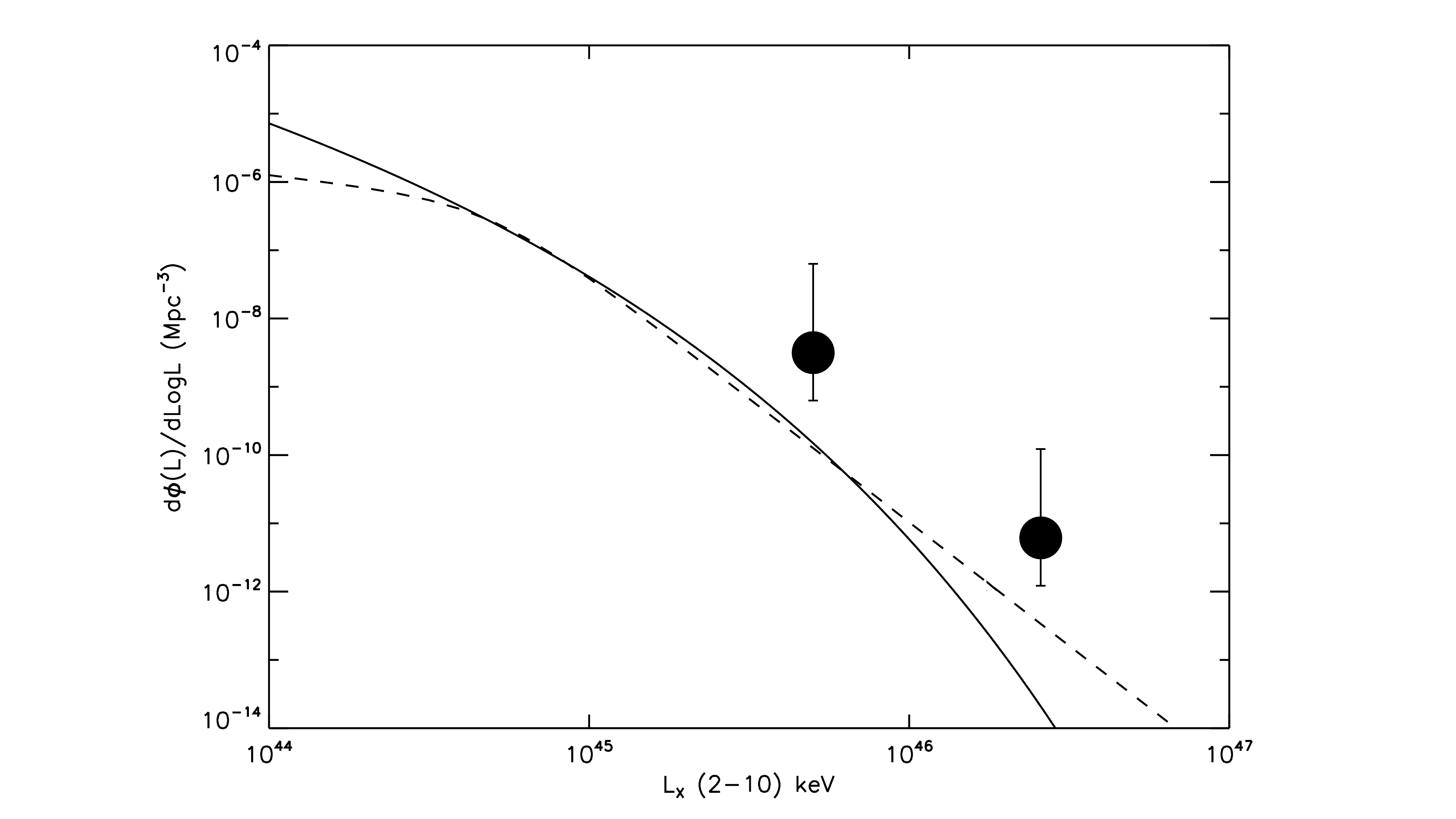}} 
\end{center}
\caption{\label{fig:qso}The black solid line represents our predicted integrated X-ray luminosity function at $z > 6$ from the early PBH-\LambdaCDM~ Universe compared with an extrapolation from the model of \citet{vito14} shown as a black dashed line. The lowest luminosity data point (with a 1-$\sigma$ error bar) is the estimate from a single detected source at $z \sim 6$ in the eROSITA eFEDS  survey reported recently by \citet{wolf}, while we derived the highest luminosity point using eROSITA data of CFHQS J142952+544717 selected from cross-correlating the Russian half of the eROSITA All-Sky Survey with the PanSTARRS Survey \citep{med}. Our prediction for the X-ray Luminosity Function is in remarkable agreement with model extrapolations and is consistent with the preliminary data points from eROSITA (as only 1-$\sigma$ errors are plotted above). }
\end{figure}

With the accretion model adopted above, we note that PBH-\LambdaCDM~ provides a recipe for the swift formation of SMBH seeds as they are naturally in place right at the time of halo assembly. In this framework, the most massive BH in the halo sinks to the center and as soon as baryons are available it starts accreting like an AGN (see schematic in Fig.~12). As the mass of the largest black hole is simply a function of the halo mass by construction in the PBH-\LambdaCDM~ model, their masses are inherently correlated (see eq. \ref{eq:npbh}). This offers a natural qualitative explanation for the origin of the locally observed $M_{\rm bh} -\sigma$ relation and accounts for co-evolution between halo assembly, galaxy assembly and BH growth. This empirical scaling relation informs all current models of the cosmic growth history of BH populations over time \citep[e.g. in][]{RicartePN2018}. The very existence and origin of such a scaling relation, whether it reflects the initial conditions of BH seed formation or is a consequence of co-evolution over time is currently debated. In the PBH-\LambdaCDM~ model, the scaling relation naturally arises out of the initial conditions and is sustained by the very nature of DM as PBHs. Since the halo is composed of PBHs, from which the central supermassive black hole originated, and depending on how we model the density profile of the halo, it will scale as $\sigma^{3-4}$. The amplitude of the relation is not predicted explicitly by the model, which is why as noted below, the \citet{bandara} value is adopted.

As the mass of the most massive BH in the halo is proportional to the mass of the halo in PBH-\LambdaCDM, we calibrate this correlation with the same relation as done by \citet{bandara} using local observational data, where $log(M_{\rm BH})=8.18+1.55*log(M_h)-13$. With the mass of the central BH settled in this manner, we impose growth at the Eddington limit and compute the emission using the template SED for an obscured SMBH as found in simulations by \citet{pacucci}. This choice is driven by the fact that at high-z the fraction of obscured AGN is essentially of order unity \citep[see e.g.][]{vito14}. However, in the case of PBH-\LambdaCDM, we do not expect all these AGN to accrete simultaneously, therefore, at each redshift, we impose the condition that the fraction of halos hosting an active AGN is given by:
$$N_{AGN}(M_h)=(\frac{M_h}{2.19\times10^{12}})^{0.9}+0.023$$
as modeled by \citet{hassan}. This number can be interpreted as either the fraction of SMBHs accreting simultaneously or alternatively as the average Eddington rate if all the SMBHs are accreting or, most likely, a combination of the two. During accretion at these early epochs, as all the AGNs are reasonably assumed to be obscured, we do not consider their contribution to reionizaton as the bulk of the UV light is absorbed in the torus/envelope. Therefore, stellar sources still contribute the bulk of the photons needed for   reionizaton and even in the PBH-\LambdaCDM~Universe replete with copious early accretors, accretion onto BHs does not significantly add to the budget. 

With these model assumptions, we find that these early AGN contribute at the level of about 2-3\% of the  CXB. In Fig.~\ref{fig:qso} we plot the predicted XLF at $z\,=\,6$ compared with data from eROSITA \citep{wolf,med} and data derived from an extrapolation of the X-ray luminosity function obtained by \citet{vito14} at z$<$5 using deep Chandra observations. Our model shows remarkable agreement with both these observational determinations, though it predicts a slight excess of sources below L$_\star$, where L$_\star\,\sim\,10^{45}\,$ergs$^{-1}$. This likely arises due to lack of data for such faint sources in observational samples that are therefore missing from the extrapolation or it could be due to our general assumptions about the accretion rate and the halo occupation statistics in the model. While further discussion of these faint sources is beyond the scope of this paper, we point out that our model provides a simple solution to the problem of SMBH seed formation without invoking hitherto unobserved physical mechanisms like extreme super-Eddington accretion or the rapid direct collapse of pristine gas.

\subsection{Stellar and AGN emission in the PBH-\LambdaCDM~ model at z$>$6.}

Next, we focus on the combined emission from star formation and AGN activity in the PBH-\LambdaCDM~ Universe. According to the model of \citet{has}, the CIB flux produced by accreting PBHs, is insufficient to account for the total CIB flux of $\sim$1 \nw needed to reproduce the observed fluctuations \citep{k2007ApJ...654L...5K,k2015ApJ...804...99K}. Additionally, \citet{helgason16} also showed that star formation at the reionizaton epoch (i.e. $z\,\sim\,{7-10}$) cannot produce the required CIB flux (see also \citet{fernandez,cooray12reio}) unless the baryon conversion efficiency is unrealistically high or lower mass halos are capable of producing stars with extremely top-heavy IMF. 

However, the PBH-\LambdaCDM~ framework could provide a further explanation. As shown above, the fraction of collapsed halos at low mass (i.e. kT$_{\rm vir}\,\sim\,10^{3-4}\,M_{\odot}$) is substantially higher than in the \LambdaCDM. This pushes star formation to earlier epochs, which goes in the right direction to account for the earlier origin of CIB. Next, we test to see if this larger abundance of lower mass halos wherein stars can form with a reasonable baryon conversion efficiency could  simultaneously satisfy all the multi-wavelength observational limits imposed by the cosmic Star Formation Rate; IR Source counts; reionizaton optical depth and and CIB versus CXB cross-fluctuations signals.

\subsection{Calculating the Ionization history} 

In order for the PBH-DM model to be viable, we need to ensure that reionizaton constraints are strictly met. The enhanced star formation at large redshifts might over-produce the ionizing photon injection rate. Using our SFR model developed in Sec.~3.1, and combining it with the predictions from the SEDs as outlined before, the injection rate can be written as:
\begin{equation}
\dot{n}(z)=\int_{13.6h^{-1} eV}^{\infty} \int \frac{J_\nu(M_h,z)}{h\nu}\,dM\,d\nu
\end{equation}
where J$_\nu(M_h,z)$ is the emissivity per unit halo mass and redshift. We estimate the IGM ionization fraction using,
\begin{equation}
\dot{x}_{ion}=\frac{f_{esc}\dot\,{n}(z) }{\langle N_H \rangle} -\frac{x_{ion}}{t_{rec}},
\end{equation}
where ${\langle N_H \rangle}$ is the mean Hydrogen column density and t$_{\rm rec}$ is the recombination time scale. From these, we can derive the Thompson optical depth via:
\begin{equation}
  \tau_e=c\sigma_T{\langle N_H \rangle}=\int x_{ion}\large(1 +\frac{\eta Y}{4X}(1+z)^2\large)\frac{dt}{dz}\,dz    
\end{equation} 
where $X$ and $Y$ are the hydrogen and helium abundance ($X$=0.76,$Y$=0.24), respectively and $\eta\,=\,1$ at $z\,>\,3$ (helium ionized once) and $\eta\,=\,2$ at $z\,<\,3$ (helium fully ionized).  In this work, we focus exclusively on sources at $z\,>\,6$, therefore in order to account for later ionization history we derive $\dot{n}_{\rm ion}$ for $z\,<\,6$ from the measured star formation histories collated in \citet{robertson}. 

\subsection{Computing Infrared source counts}

Having developed a model for growth of structure and emission, we can ask ourselves how many of these faint early sources in PBH-\LambdaCDM~ we can expect to detect in IR surveys. Accretion onto PBHs produces very faint X-ray emission, so we estimate the number counts of proto-halos/galaxies by integrating the CLF in the IR. First we derive the luminosity function:
\begin{equation}
\phi(L_{\nu},z)=\int_{M_{Min}}^{M_{Max}} \phi(L_{\nu}(M_h,z))\,dM;
\end{equation}
from which we then obtain the number counts,
\begin{equation}
N(m)=\int_0^\infty \phi(m_{\nu},z)\,dV(z)\,dz
\end{equation}
where $dV(z)$ is the co-moving volume element at redshift $z$. Observed source counts from the soon to be launched James Webb Space Telescope (JWST), in the the [2-5] $\mu$m~band with NIRCAM offer the prospect for placing tight constraints on our model as reported in the Results section of this paper.

\subsection{The production rate of Cosmic Backgrounds}

In order to evaluate the integrated cosmic backgrounds produced by these additional high-z mini-halo populations in the PBH-\LambdaCDM~ Universe, we first need to compute the emissivity of individual star forming halos for which we deploy spectral templates from the {\it Yggdrasil} model \citep{z11}.{\it Yggdrasil} creates synthetic spectra that include a single age stellar population and a mix of Pop II and Pop III stars. The parameters of the model are simply the Initial Mass Function (IMF) and metallicity. With these parameters the spectra are modeled to include nebular emission and  extinction/reprocessing from dust. As done in \citet{helgason16}, we assign different IMFs to the halos according to their masses: for halos with $T_{vir}\,>\,4\,\times\,10^{4}$ K, and if they are already enriched, we adopt a Pop II Kroupa IMF in the mass range 0.1-10 M$_{\odot}$ with a metallicity $Z\,=\,0.0004$ (hereafter, our Model IMF2). The other free parameters of the model are the escape fraction $f_{esc}$ and the star forming efficiency floor f$_{\star_{\rm min}}$. We refer to \citet{helgason16} and \citet{z11} for a detailed description of the spectral templates.  Cooling in the lower mass halo population  is driven by H$_2$ cooling, so for these halos, we assume that the IMF is the same as assumed for the formation of Pop III stars. To model the IMFs in these lower mass mini-halos, we explore three zero metallicity IMFs, a Top-Heavy IMF with a cut-off at 500 M$_\odot$ (IMF-3A), a lognormal  with a characteristic mass of 10 M$_\odot$ and $\sigma_M$= 1 (IMF-3B) and a Kroupa IMF with a cut-off at 100 M$_\odot$ (IMF-3C), models referred to as IMF-3A, IMF-3B and IMF-3C respectively. For each SED, the emissivity per unit DM halo mass M$_h$ per unit redshift is given by:
\begin{equation}
j_\nu(M_h,z)=\int_z^{\infty} \dot{\rho_{\star}}(M_h,z^\prime)L_\nu(t-t_z^\prime)\frac{dt}{dz^\prime}\,dz^\prime,
\end{equation}
where $L_\nu(t-t_z^\prime)$ is derived from the evolving spectral templates from {\it Yggdrasil}. For a given wavelength, we can now derive the flux production rate, 
\begin{equation}
f(M_h,z)=\frac{c}{4\pi}j_\nu(M_h,z)\frac{\nu}{1+z}\frac{dt}{dz}.
\end{equation}
We now have the $f(M_h,z)$ necessary for Eqn.~\ref{eq:power} to derive $P_2(q)$. Similarly, we can derive the CLF by deriving the halo luminosity:
\begin{equation}
L(M_h,z)=\frac{\nu J_\nu(M_h,z)}{(M_hdn/dM_h(z))}
\end{equation}
from which we obtain that: $\phi(L(M_h,z))=\frac{dN}{dM_h(z)}$.
Since {\it Yggdrasil} produces only results for rest-frame wavelengths longer than the UV, we derive the X-ray emissivity and CXB production rate using an extrapolation of the L$_X$-SFR relation of \citet{aird}:
\begin{eqnarray}
log(L_X)=39.48+0.83\,log(SFR)+1.34\,log(1+z)
\end{eqnarray}
and convert it into flux. This allows us to compute the CXB and CIB produced by this source population in the PBH-\LambdaCDM~ Universe.

\subsection{Calculation of fluctuations in the cosmic backgrounds}

With the PBH-DM halo populations and their emissivity from star formation and AGN activity computed, we then use an appropriately modified version of the HOD formalism to compute the integrated surface brightness and fluctuations therein. At a given frequency $\nu$, we define the surface brightness fluctuations of the diffuse component of background radiation as 
\begin{eqnarray}
\delta F_{\nu}(x)=F_{\nu}(x)-\langle F_{\nu} \rangle,
\end{eqnarray}
where F$_\nu$ is the surface brightness at the position $x$ in the sky and $\langle F_{\nu} \rangle$ is the mean integrated flux of the background at the frequency $\nu$. We evaluate the power spectrum of fluctuations defined as: $P(q)=\langle|\delta{\bf F_{\nu}}(q)|\rangle$ with $q$ as the angular wave number and where $\delta{\bf F_{\nu}}(q)$ is the Fourier transform of the 2D fluctuation field. Here, we adopt the formalism derived by \citet{ricarte19}. The power spectrum can be written as a sum of the one- and two-halo terms:
\begin{equation}
    P(k) = P^{1h}(k) + P^{2h}(k).
\end{equation}
These two components represent contributions to flux from within and from clustering outside the halo respectively. For each source population (see below) we define
\begin{equation}
    F^c_\nu = \langle f^c_\nu (M_h,z) \rangle  ,
\end{equation}

\noindent and

\begin{equation}
    F^s_\nu = \langle f^s_\nu(M_h,z) \rangle .
\end{equation}
 where f$^c_\nu$ and f$^s_\nu$ are background production rates for each source population of PBHs within halos of given mass in central (denoted by the super-script c) and satellite subhalos (denoted by the superscript s) at a frequency $\nu$, respectively. For the populations described here, we assume that all PBH sources, even the ones that constitute the parent halo are essentially satellites except for the single most massive central BH that accretes like an AGN. We assume that PBHs are distributed within individual halos in accordance with the expected Navarro-Frenk-White profile \citep[NFW][]{nfw} predicted by CDM. The background production rate in a given M$_h$ bin and redshift is described in Sec.~4.2. As each population traces dark matter differently, their two halo terms are then given by:
 \begin{eqnarray}
 P^{2h}_{i,j}(k,z)\,=\,b(M_h, z)_i\,*\,b(M_h, z)_j\,*P^\mathrm{lin}(k,z),
 \end{eqnarray}
 where $P^\mathrm{lin} (k,z)$ is PBH-\LambdaCDM~ power spectrum from Eqn.~1 and $b(M_h, z)_i$ is the linear halo bias \citep{Sheth+2001}, for the i-th population. As we are dealing with diffuse emission rather than individual point sources, bias is now computed by flux-weighting central and satellite emissivities such that:

\begin{equation}
    B^c = \int \frac{dn}{dM_h} \frac{F^c_\nu}{\bar{F_\nu}} b(M_h,z) dM_h
\end{equation}

\noindent and

\begin{equation}
    B^s = \int \frac{dn}{dM_h} \frac{F^s_\nu}{\bar{F_\nu}} b(M_h,z) u(k|M_h,z) dM_h
\end{equation}

\noindent where  $u(k|M_h,z)$ is the Fourier transform of the NFW profile.  We can then write out the 1-halo term as

\begin{align}
    & P_{i,j}^{1h}(k,z) = \int \frac{dn}{dM_h} \times \\ 
    &\nonumber \frac{(F^s_iF^c_j + F^s_jF^c_i)u(k|M_h,z) + F^s_iF^s_ju^2(k|M_h,z)}{\bar{F}_i \bar{F}_j} dM_h ,
\end{align}

 and the two-halo term as:

\begin{equation}
    P^{2h}_{i,j}(k,z) = P^\mathrm{lin}(k,z)(B^c_i + B^s_i)(B^c_j + B^s_j) ,
\end{equation}

The indices $i,j$ label the photometric bands so that if $i=j$ we obtain the auto-power spectrum in a single band while, if  i$\neq$j one obtains the cross-power spectrum. Next, we reconstruct the power spectrum using Limber's equation that projects the 3D power spectrum folded in with the emissivity of each halo: \\
\begin{equation}
\begin{split}
P_2(q)=\int \frac{H(z)}{cd^2(z)}
        \int f_{\nu_{i}}(M_h,z)\,f_{\nu_{j}}(M_h,z) \\ P_{3_{i,j}}(qd_c(z)^{-1})\,dM\,dz,
    \label{eq:power}
\end{split}
\end{equation}
\\
where, $q$ is the angular wave-number in rad$^{-1}$; d$_c$(z) is the comoving distance; $H(z)$ is the expansion rate of the Universe at redshift $z$; $f_{\nu_{i,j}}(M_h,z)$ is the flux produced per unit halo mass and redshift at a frequency $\nu_{i,j}$; and P$_{3_{i,j}}(qd_c(z)^{-1})$ is the dark matter power spectrum described in Eqn.~\ref{eq:PBH} expressed a function of $q$. We now have all the requisite machinery in place to estimate the unresolved background fluctuations in the X-ray [0.5-2] keV and [2-5] $\mu$m bands, their auto- and cross-power spectra. 
 \\ 

\subsection{Auto and Cross-correlations of cosmic backgrounds}

A powerful set of observables predicted by our model are the CIB and CXB fluctuation power spectra and their cross-power. We have computed these by including the flux production rate in Eqns.~5 \& 6 by assuming that stars and the general population of growing PBHs are distributed as satellites and that the central sources - the most massive PBHs in their parent haloes - grow like an AGN. Our final angular power spectra are then obtained as follows:
\begin{equation}
P_{i,j}(q)=P_{i,j}^{Stars}(q)+    P_{i,j}^{QSO}(q) + P_{i,j}^{PBH}(q)  + XP_{i,j}(q)+ P^{SN}
\end{equation}
we enclose in $XP(q)$ the cross power terms for each pair of populations and bands. 

We have added as foregrounds shot-noise and clustering from unresolved {\it local} galaxies, AGN and clustering both in the IR and X-ray bands using the predictions of \citet{helgason12,helgason14} who extrapolated known luminosity functions of galaxies, AGN and clusters to fluxes below current survey flux limits. Here, we have assumed a fiducial flux limit in the X-ray of S$_{lim,0.5-2}=10^{-17}$ erg cm$^{-2}$ s$^{-1}$ in the [0.5-2] keV band and $m_{\rm AB}\,=\,24.5$ in the [2-5] $\mu$m band. 

\section{Results: Observational Signatures of a PBH-\LambdaCDM~ Universe}

As with all treatments that involve extrapolating current observations to higher redshifts, to $z\,>\,7$, be it in the context of the standard \LambdaCDM~ model or variants like the PBH-\LambdaCDM~ that we consider here, assumptions need to be made regarding astrophysical processes relevant to star formation, black hole accretion and the corresponding emitted flux at these extremely early epochs. As a result, unsurprisingly, we find that there exist classes of PBH-\LambdaCDM~ models that are viable and consistent with current observational constraints. Our first key finding, modulo our assumptions detailed in Sec.~3., is that there are two parameters that drive the best-fits for the class of feasible models and they are: f$_\star$ - the efficiency floor for star formation and the $f_{\rm esc}$ the escape fraction of ionizing photons; quantities that are empirically ill-constrained even at much later cosmic epochs.

In order to determine the optimal combination of parameters for our models that are in agreement with observed constraints, we sampled the parameter space using the {\em Python} package {\em emcee} \citep{emcee} that uses the affine-invariant ensemble sampler for Markov chain Monte Carlo (MCMC). We run these chains assuming flat prior conditions on all our parameters (i.e. all the values sampled have the same prior probability). The priors for our models are listed in Tab. \ref{tab:results} in the form of intervals outside which the Likelihood become -$\infty$.

To create our MCMC chains we employed 30 walkers and 1000 steps, from which we then define a likelihood function to determine the posterior probability in the following usual form:
\begin{equation}
\begin{split}
\ln{P}= -\frac{1}{2}+ \frac{(\tau-\tau_{m})^2}{\sigma_{\tau}} +
\frac{(SFR-SFR_{m})^2}{\sigma_{SFR}},
\end{split}
\end{equation}
where the index $m$ represents the m-th realization of the model; $\tau$ is the Thompson optical depth measured by \citet{planck}; SFR is the star formation density at $z\,=\,7$ estimated by \citet{Bouwens}. Additionally we impose the condition that the total X-ray background CXB produced by our sources in the PBH-\LambdaCDM~ does not exceed the unresolved component in the [0.5-2] keV band constrained by \citet{cap17}. In the event that a model exceeds the observationally constrained CXB level, we set $\ln{P}\,=\,-\infty$. We then repeat the sampling for each combination of choice of Pop III IMFs and enrichment history with the following models: IMF-3A (Top-heavy Pop III IMF with a cut-off at 500 M$_{\odot}$ ), IMF-3B (Lognormal Pop III IMF with a characteristic mass of M $\sim$ 10 M$_{\odot}$), IMF2 (Pop II Kroupa IMF in the mass range 0.1-10 M$_{\odot}$ with a metallicity $Z\,=\,0.0004$ ) and IMF-3C (Pop III Kroupa IMF with a cut-off of 100 M$_{\odot}$). The free parameters of the overall model are: f$_\star$ and f$_{\rm esc}$ for each population - $\eta_{III}$, f$_{\rm esc,III}$, $\eta_{II}$ and f$_{\rm esc,II}$ for the Pop III and and Pop II halos, respectively. We ran independent fits for each combination of choice of Pop III IMF template and enrichment history, for a total of 9 explorations. The best fit results for each combination are summarized in Table \ref{tab:results}. The posterior distributions for our MCMC fits are shown in the Appendix in Fig.~\ref{fig:app}. 

Overall, since the fit relies on and is driven only by two parameters, several scenarios are feasible and can reproduce the observed Universe and are comparable with current observational constraints. Our results provide clarity on the possible permitted scenarios. Now, we focus on key predictions of the model scenarios that are viable and consistent with the current scant, high redshift observational constraints. From Figs.~4-11, it is clear that the fits provide significant constraints for each parameter with the notable exception of the late enrichment scenario, in which $\eta_{III}$ and f$_{\rm esc,III}$ are degenerate. The region with the parameter space at high  $\eta_{III}$ and high f$_{\rm esc,III}$ can be excluded conclusively by our model, likely due to the fact the such a combination would produce too many ionizing photons. For $\eta_{II}$  and f$_{\rm esc,II}$ our fits point to a feasible scenario where the star formation efficiency is of the order of a fraction of a percent and the escape fraction is significantly lower than 10\%. By construction the PBH-\LambdaCDM~ Universe explored here patches on to the \LambdaCDM~ Universe at $z\,\leq\,6$. In order to keep the model economical we do not include a halo mass dependent escape fraction, and therefore these reported values must be considered as an average for the population. While constraining the properties of early star formation is an open question in astrophysics, addressing it any further other than with these simple parametrizations is beyond the scope of this work. Our goal has been to to explore the possible permitted realizations of an early Universe within a simple PBH-\LambdaCDM~ cosmology that has explanatory power.
\
\begin{table}[]
    \centering
    \begin{tabular}{|l|c|c|c|l|}
\hline
  IMF-3A   & Early & Mid & Late & Priors \\
  \hline
  \hline
  Log($\eta_{III}$) &-3.41$_{-1.07}^{+1.44}$&   -3.63$_{-0.91}^{+1.40}$ &  -3.34$_{-1.19}^{+1.56}$ & (-5,-1)\\
Log(f$_{escIII}$) & -2.61$_{-0.90}^{+1.62}$ & -2.58$_{-0.98}^{+1.49}$ &  -2.41$_{-1.09}^{+0.97}$&(-4,0) \\
Log(f$_{escII}$) &  -2.23$_{-0.99}^{+0.34}$& -2.27$_{-0.64}^{+0.33}$  &-2.79$_{-0.83}^{+0.75}$ &(-3.5,-1)\\
Log($\eta_{II}$) & -2.22$_{-0.28}^{+0.22}$ & -2.21$_{-0.23}^{+0.17}$ &   -2.17$_{-0.47}^{+0.20}$ & (-4,0) \\
    \hline
     \hline
     IMF-3B & & \\
        \hline
        \hline
   Log($\eta_{III}$)    & -3.46$_{-1.09}^{+1.44}$&   -2.94$_{-1.33}^{+1.22}$ &   -3.21$_{-1.22}^{+1.44}$ & (-5,-1) \\    
Log(f$_{escIII}$)       &  -2.43$_{-0.77}^{+1.04}$ &   -2.43$_{-0.77}^{+1.04}$ & -2.55$_{-0.88}^{+1.14}$  &(-4,0)\\
    Log(f$_{escII}$)     &   -2.11$_{-0.37}^{+0.32}$  &   -2.22$_{-0.97}^{+0.37}$& -2.26$_{-0.64}^{+0.45}$ &(-3.5,-1) \\
   Log($\eta_{II}$)     &   -2.24$_{-0.38}^{+0.21}$  &  -2.24$_{-0.30}^{+0.19}$&   -2.25$_{-0.65}^{+0.27}$ & (-4,0)\\
   \hline
     \hline
     IMF-3C & & \\
         \hline
        \hline
   Log($\eta_{III}$)    & -3.14$_{-1.19}^{+1.43}$ & -3.01$_{-1.25}^{+1.11}$  & -3.17$_{-1.24}^{+1.50}$ & (-5,-1)  \\    
Log(f$_{escIII}$)     & -2.40$_{-1.11}^{+1.60}$ &  -2.43$_{-1.03}^{+1.3}$  & -2.12$_{-1.19}^{+1.44}$  &(-4,0) \\
 Log(f$_{escII}$)    & -2.10$_{-0.34}^{+0.26}$  &  -2.06$_{-0.33}^{+0.26}$  & -2.19$_{-0.73}^{+0.34}$  &(-3.5,-1)  \\
  Log($\eta_{II}$)    & -2.24$_{-0.33}^{+0.18}$   & -2.28$_{-0.24}^{+0.21}$ &  -2.20$_{-0.44}^{+0.25}$ & (-4,0)  \\
  \hline
     \hline

    \end{tabular}
    \caption{Best fit results for our MCMC realizations for the 3 assumed enrichment histories and the 3 assumed Pop III IMFs.}
    \label{tab:results}
\end{table}

\subsection{Detailed model predictions}

\subsubsection{Star formation rate density and accreted mass density}
\begin{figure*}[!t]
\center
\includegraphics[width=0.73\textwidth]{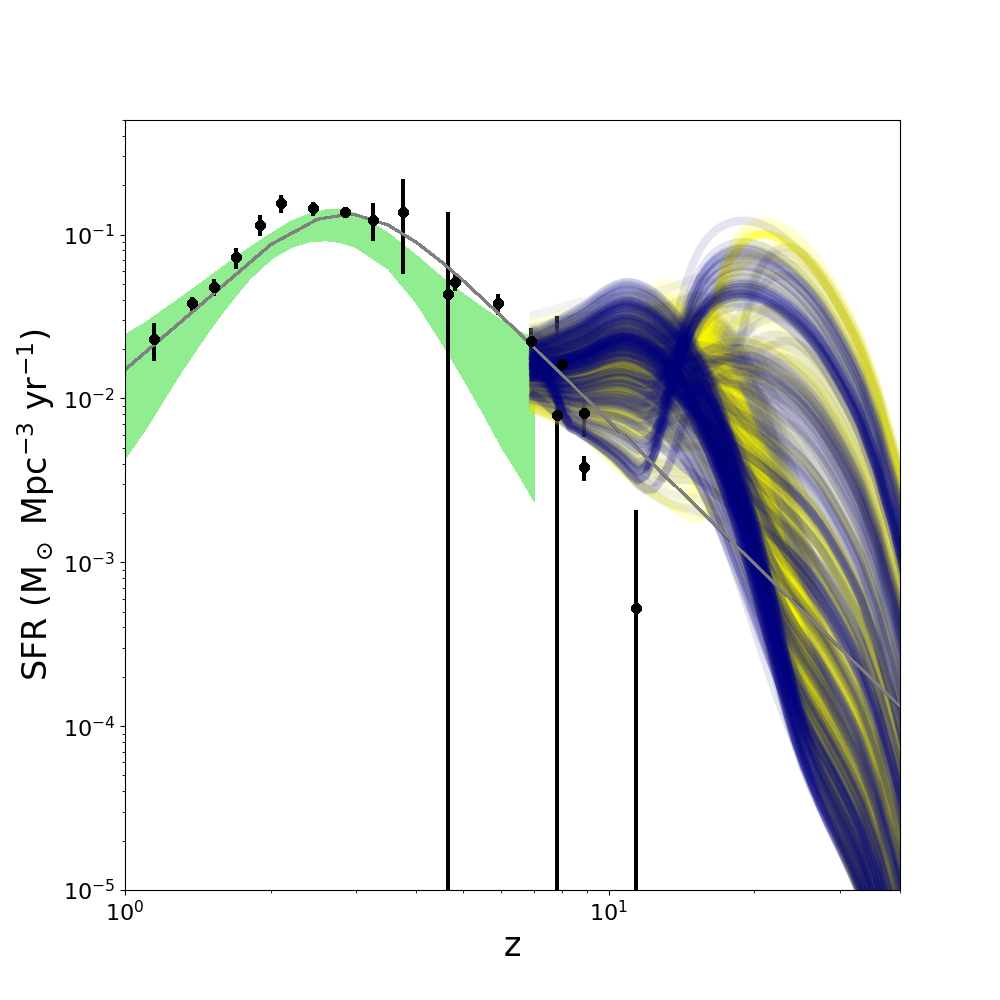}
\caption{\label{fig:sfr} The Star Formation Rate Density (SFRD) for our model realizations. The grey, yellow and navy lines represent {\bf Early, Medium and Late} enrichment, respectively. Our $z > 7$ model tracks are compared with "local" measurements from either EBL \citet{ajello} or high-z surveys \citep[see e.g.][]{Bouwens} shown as the green band. The grey continuous line and black data points represent the best fit performed and the data collected by \citet{madau} and refs. therein. The density of the lines represents the sampling probability of the parameter space. In our fits, the SFRD is driven by $\eta_{II,III}$ and their best fit values are consistent, and the star formation rates predicted are independent of the IMF choice for Pop III stars. Therefore, here we only show the realizations of the model for IMF-3A.} 
\end{figure*}  
In Fig.~\ref{fig:sfr} we show the predicted star formation rate density as function of redshift for all realizations with a 1$\sigma$ dispersion as drawn from our fits. In the main text we show the case for IMF-3A while the other three cases IMF-2, IMF-3B and IMF-3C are reported in the Appendix. We also show the three cases with different enrichment histories - the {\bf Early, Medium and Late} considered here that all smoothly connect with the current SFRD plotted in green, determined from observations of EBL \citep{ajello} and high-z galaxy surveys \citep[best fit and data from][]{madau}. We note that by construction, all model variants are tuned to reproduce the z$\sim$6-7 star formation rate density, therefore our predictions are primarily for $z\,>\,7$. The density of model lines plotted represents the model sampling density. As can be seen, while all the three IMFs produce a similar dispersion overall, the specific track families predicted for each enrichment model differ. The generic prediction is the existence of two clear peaks in the SFRD between $10\,<\,z\,<\,30$. Each enrichment model has a distinct peak. The realizations suggest that in the case of early enrichment the bulk of the star formation happens in mini halos at very high-z and Pop III halos become dominant at z$<$10. In the late enrichment case, on the other hand, most of the Pop III activity is delayed and therefore it dominates the re-ionzation epoch.

\subsubsection{Thompson Optical Depth}

\begin{figure*}[!t]
\center
\includegraphics[width=0.71\textwidth]{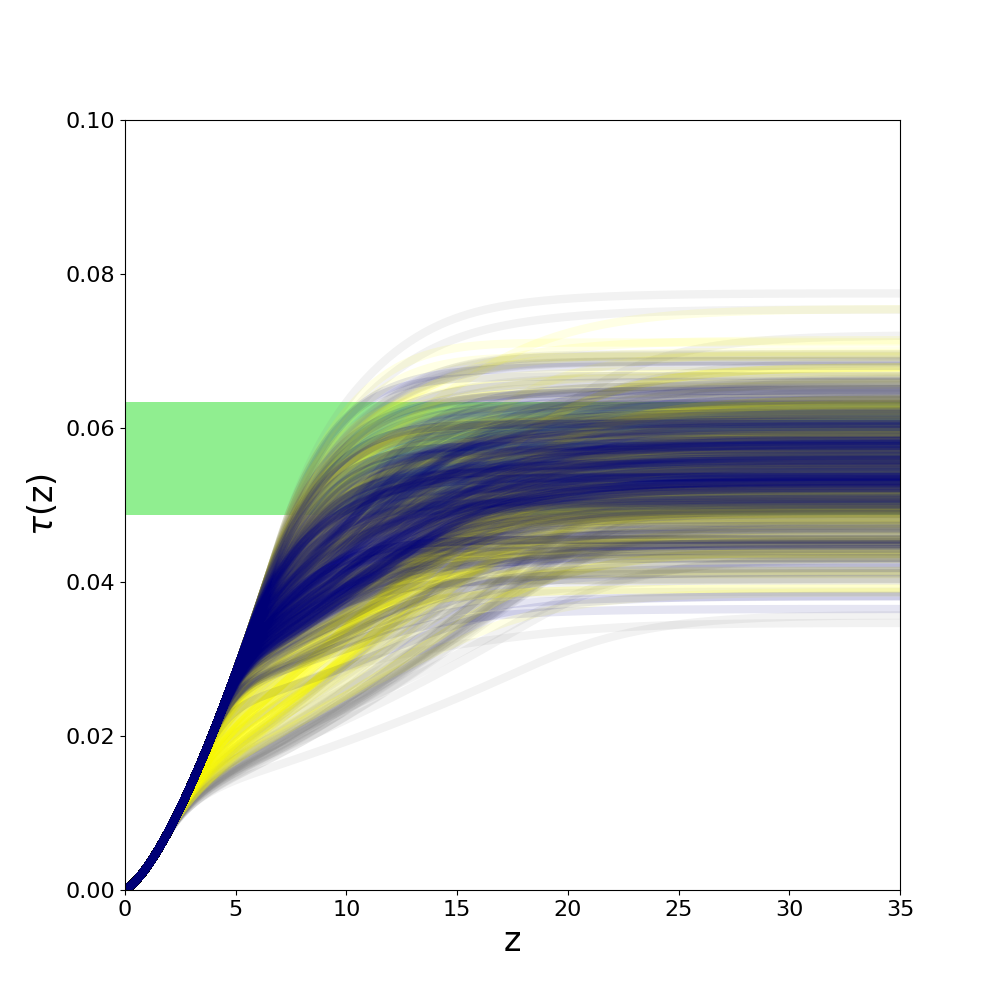}
\caption{\label{fig:CMB} The redshift evolution of the Thompson optical depth compared with Planck limits \citep{planck} represented as a green band. This figure reports results for the IMF-3A model. The grey, yellow and navy lines represent {\bf Early, Medium and Late} enrichment, respectively. The density of the lines represents the probability of realization of our model based on the sampling of the parameter space. Our model predicts several compatible reionization history scenarios for various combinations of f$_\star$ and f$_{\rm esc}$.}
\end{figure*}

For the assumed SF history in the PBH-\LambdaCDM~ model, we can compute the predicted redshift evolution of the Thomson optical depth for the three enrichment models and for the class of explored IMFs. For IMF-3A we show these in Fig.\ref{fig:CMB}, where current constraints from Planck \citep{planck} are shown as a band in green. As noted above, many classes of models with various combinations of values of $f_{\rm esc}$ and $f_*$ considered here are, in principle, consistent with current observations.

\subsubsection{CIB and CXB flux density}
 
With the emission from star formation and BH accretion modeled as described in Secs.~3.3 and 3.4; we calculate the cumulative flux densities for the resultant CIB and CXB from the entire population at $z\,>\,7$, as outlined in Sec.~3.7. The contribution from accreting AGN is shown with the black dashed line and that from Bondi accretion onto PBHs is shown with the black dotted line in Fig.\ref{fig:CIB} for the model IMF-3A. The other two Pop III IMF cases, IMF-3B and IMF-3C are reported in the Appendix. The clear picture that emerges from these PBH DM models is that the bulk of the flux contribution to the CIB is provided by star formation activity and accretion processes are a sub-dominant component.
All the model predictions produce $<1$ nW m$^{-2}$ sr$^{-1}$ of the CIB. This value is what is generally assumed to be necessary to safely account for the observed fluctuations. However, our model does not rule out a scenario in which first light commences with an early highly efficient star formation episode, followed by a less efficient Pop II episode. In this scenario, the escape fraction tends to be extremely low in Pop III stars, of the order 10$^{-3}$ suggesting that such an episode must occur in an optically thick environment from the point of view of UV light production. 

 \begin{figure*}[!t]
 \center
\includegraphics[width=0.71\textwidth]{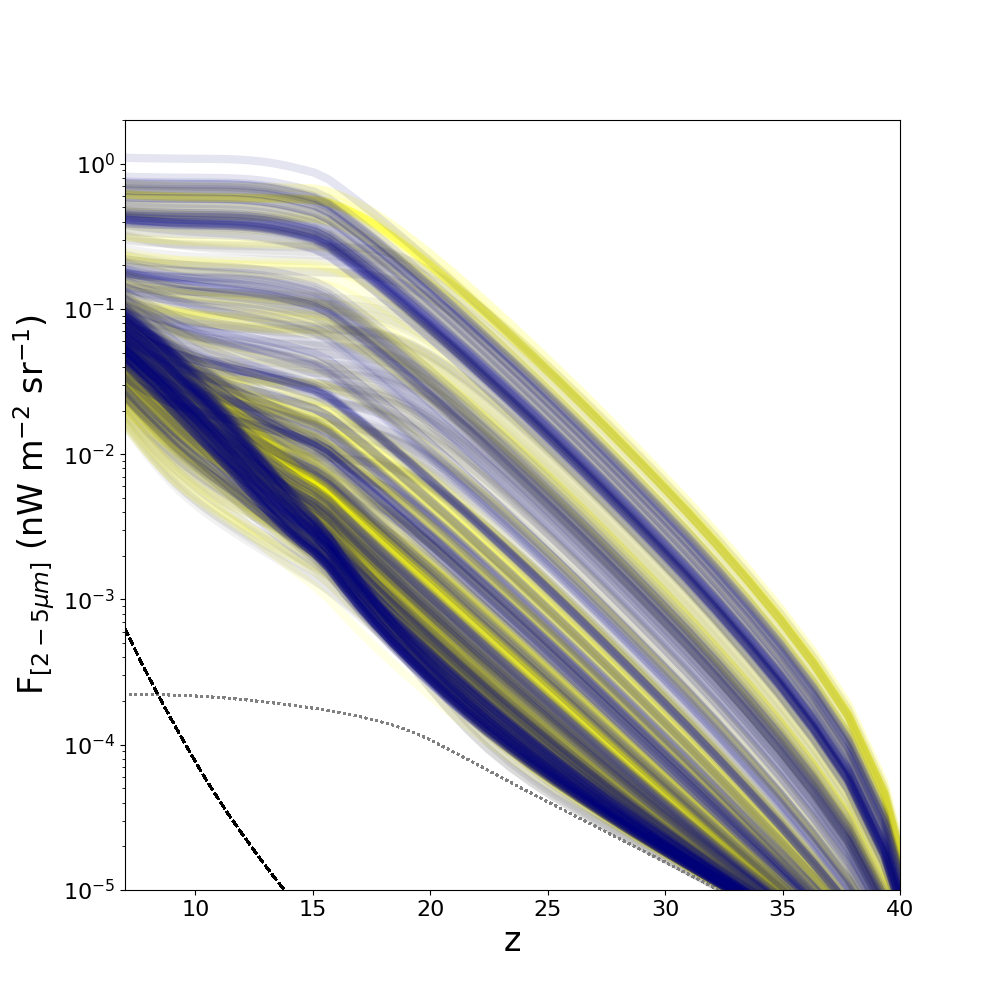}
\caption{\label{fig:CIB} CIB Flux production rate as a function of redshift plotted for realizations of the IMF-3A model. The grey, yellow and navy lines represent {\bf Early, Medium and Late} enrichment, respectively. The black dashed line represents radiation produced by AGN while black dotted line represents accretion from streaming baryons onto PBHs. The density of the lines represents the probability of realization of our model based on the sampling of the parameter space. The bulk of the CIB is produced by star formation while AGN are subdominant for this census.}
\end{figure*}

\begin{figure*}[!t]
\center
\includegraphics[width=0.7\textwidth]{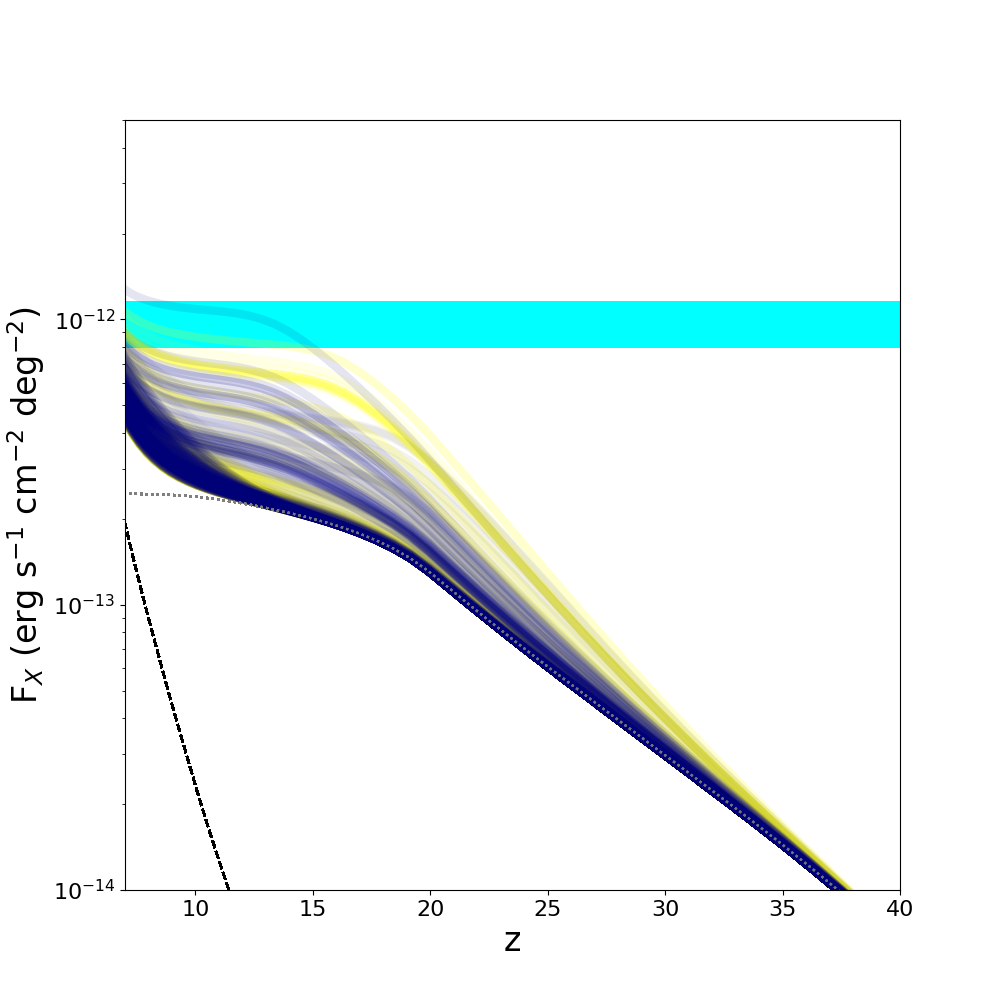}
\caption{\label{fig:CXB} The cumulative CXB flux production rate as a function of redshift from realizations of the IMF-3A model. The grey, yellow and navy lines represent {\bf Early, Medium and Late} enrichment, respectively. The black dashed line represents radiation produced by AGNs while black dotted line represents accretion from streaming baryons onto PBHs. The density of the lines represent the probability of realization of our model based on the sampling of the parameter space. The $Cyan$ band represents the current limit on the unresolved CXB from \citet{cap17} and \citet{hm06}. At z$>$15 the bulk of the CXB is produced by PBH accretion while, at z$<$15 AGN become the dominant sources. X-ray binaries contribute at the few percent level to the CXB.}
\end{figure*}

Similarly, we compute the contribution to the CXB from these $z\,>\,7$ sources, and these are shown in Fig.~\ref{fig:CXB}, where the black dashed line again shows the contribution from AGN and the black dotted line the contribution from streaming baryons. The current limits on the unresolved component of the CXB is shown as the cyan band. At $z\,>\,15$ the bulk of the CXB is produced by PBH accretion while, at $z\,<\,15$ AGN hosted at the centers of PBH DM halos become the dominant contributor. 

\subsubsection{Source Counts}

\begin{figure*}[!t]
\center
\includegraphics[width=0.7\textwidth]{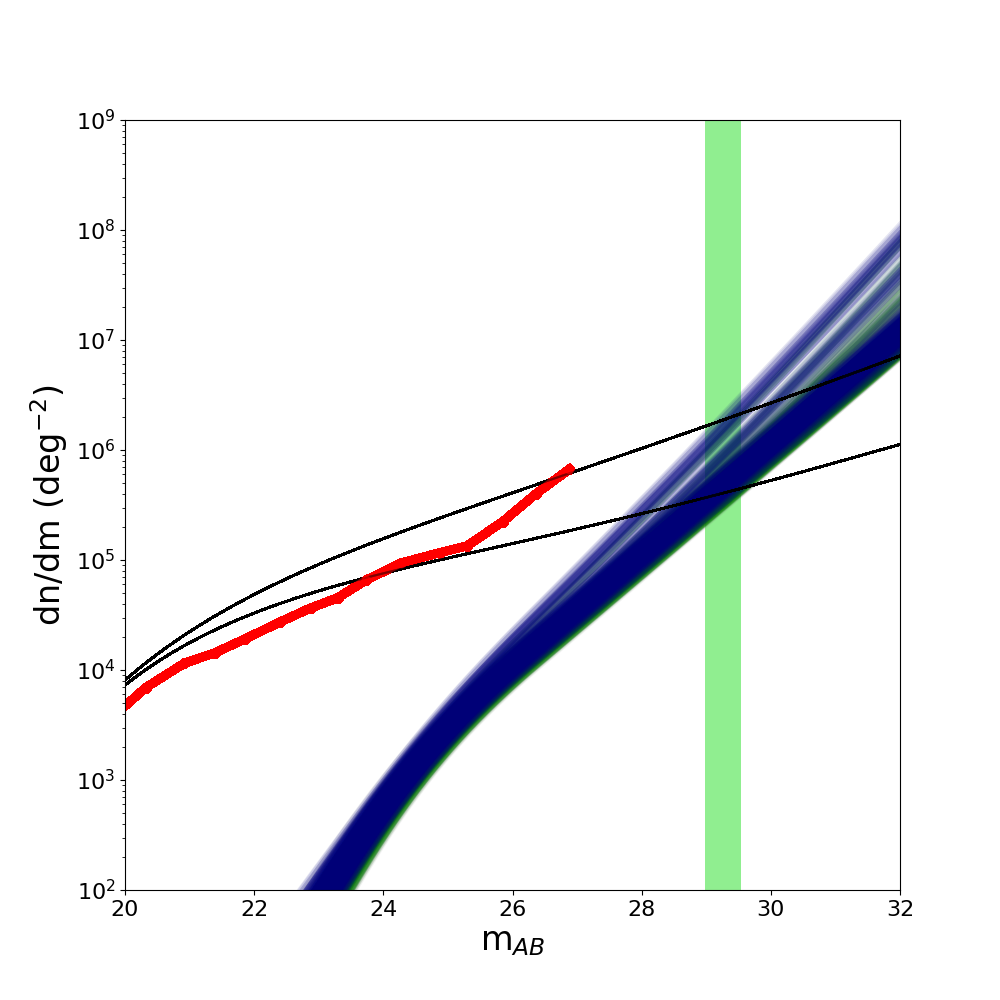}
\caption{\label{fig:counts} The predicted source counts for $z\,>\,6$ NIR sources from realizations of the IMF-3A model. The grey, yellow and navy lines represent {\bf Early, Medium and Late} enrichment, respectively. The predictions are compared with S-CANDLES data from \citet{2015ApJS..218...33A} (red line) and with  extrapolations of the population synthesis model of \citet{helgason12} based on the faint end of the luminosity function. All the models predict comparable results with a steepening of the counts at $m_{\rm AB} > 28-30$ with a slightly earlier onset of the early population in IMF-3A (see Appendix for comparison with predictions of the IMF-3B and IMF-3C models). The Green band represents the JWST 10 ks magnitude limit for 10ks exposure.}  
\end{figure*}

Applying appropriate bolometric corrections, we compute the predicted NIR source counts. At faint magnitudes $m_{\rm AB} \sim 28-30$ the contributions from PBH-\LambdaCDM~ sources completely dominates. In Fig.~\ref{fig:counts}, we show our model predictions along with high redshift extrapolations of the faint end of current observed IR luminosity functions derived using population synthesis models produced by \cite{helgason12}. It turns out that these computed faint IR source counts offer the most stringent test of the PBH-\LambdaCDM~ model. Observations from NIRCAM aboard JWST are expected to detect sources in precisely this magnitude range $m_{\rm AB} \sim 28-30$ where our models clearly predict a significant excess in counts as well as an unambiguous, strong steepening of the number count slope. 

In Fig.~\ref{fig:counts} we compare the deep S-CANDELS results  \citep{2015ApJS..218...33A} with a prediction of the 4.5 $\mu$m number counts from our model for the Pop III IMFs considered here and the three enrichment histories and assuming f$_\star$=0.005 and an escape fraction f$_{\rm esc}$=0.1. The predictions are compared with the extrapolations of the population synthesis model of \citet{helgason12} based on the faint end of the luminosity function. Star formation in the PBH-\LambdaCDM~ model does not produce enough flux to significantly change the shape of the source counts at $m_{\rm IRAC}\,<\,{28-30}$, but at the very faint end (i.e. $m\,\gtrsim\,{28-30}$) we predict, that due to the rapidly steepening $\log N - m$ relation, early star forming objects become the dominant population. Interestingly, this  steepening lies right at the expected depth of the JWST 10 ks survey like the JWST-NEP \citep{2018PASP..130l4001J}. Therefore, this important model prediction stands to be tested very soon.
   
\subsubsection{CIB and CXB angular auto and cross-power spectra}

\begin{figure*}[!t]
\center
\includegraphics[width=0.7\textwidth]{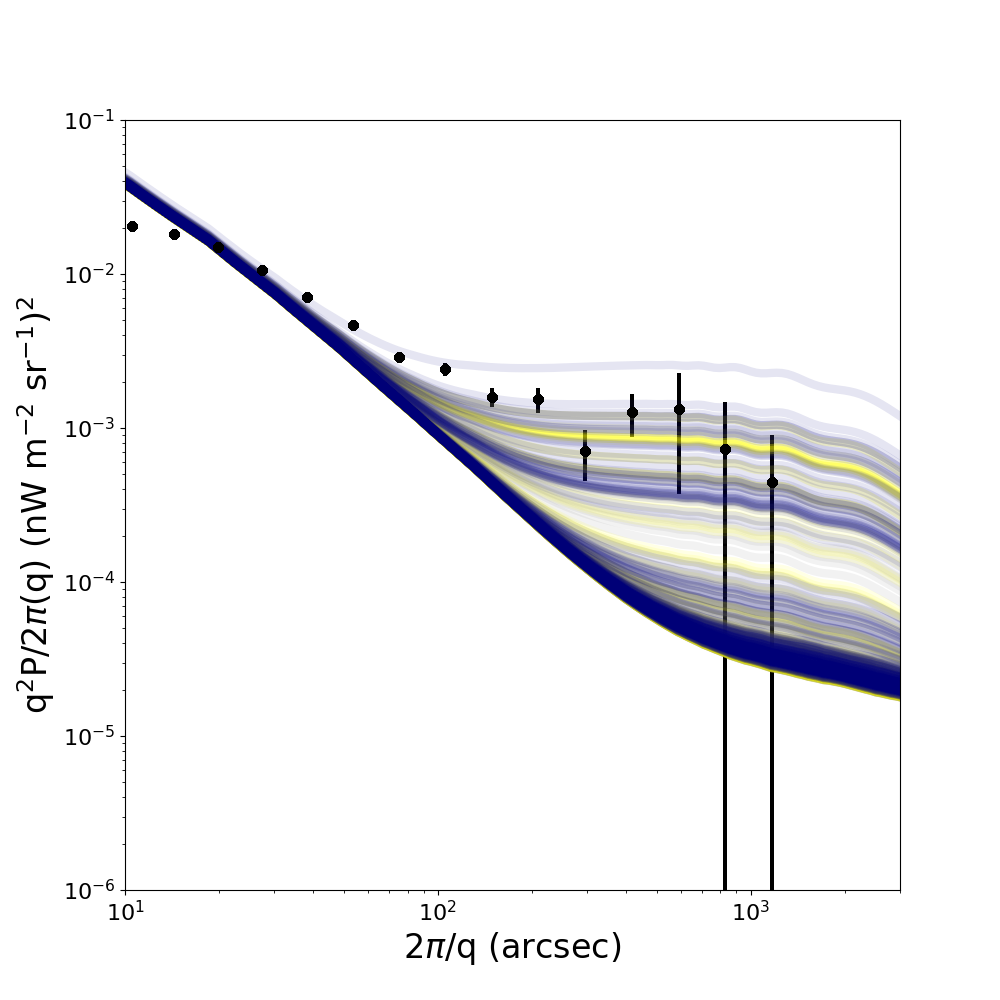}
\caption{\label{fig:CIBPS} The power spectrum of the unresolved [2-5]$\mu$m CIB fluctuations from realization of our model IMF-3A added to the foreground galaxy estimate from \citet{helgason12}. Data points are  the weighted average of the  results of \citet{2012ApJ...753...63K,li18}. The grey, yellow and navy lines represent {\bf Early, Medium and Late} enrichment, respectively.  The density of the lines represents the probability of realization of our model based on the sampling of the parameter space. Very few realizations with the late enrichment model can explain the excess observed over the foreground signal.} 
\end{figure*}

\begin{figure*}[!t]
\center
\includegraphics[width=0.7\textwidth]{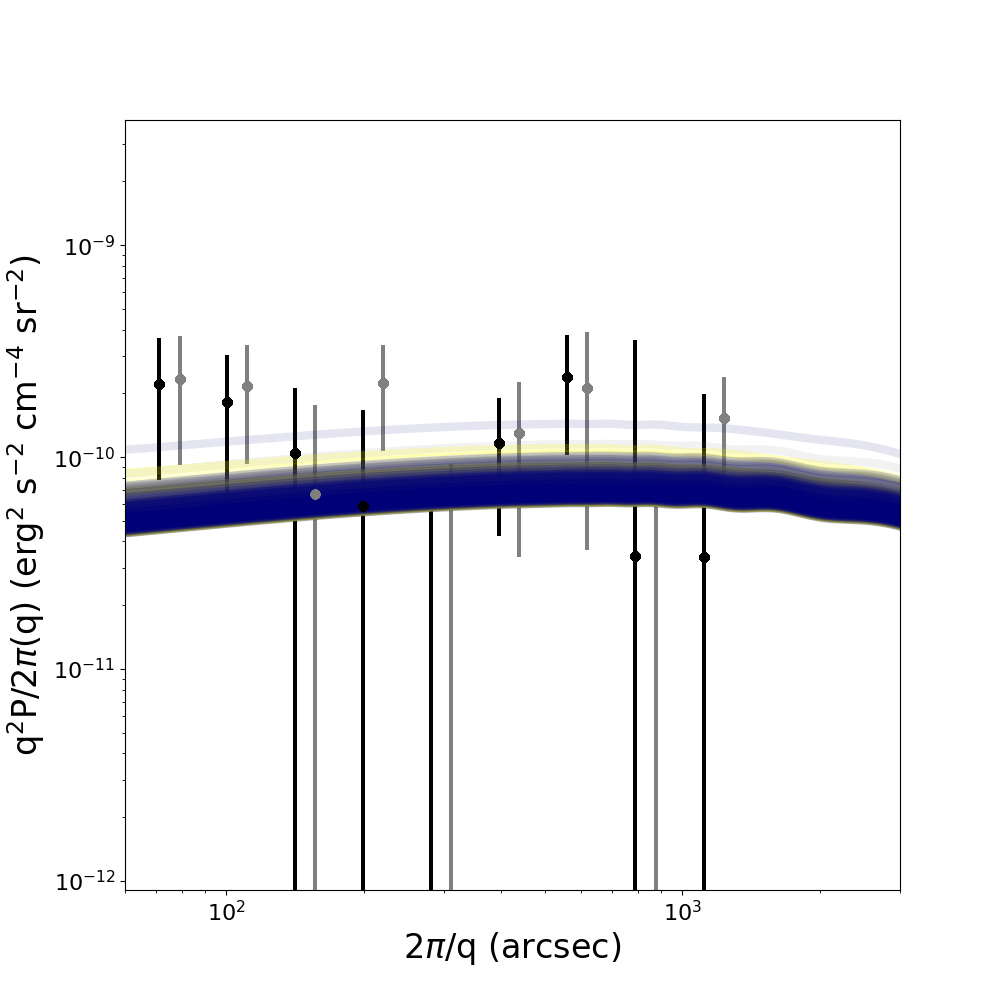}
\caption{ \label{fig:XPS} The power spectrum of [0.5-2] keV sources: computed from the IMF-3A model, these include the contributions from accreting PBHs and X-ray Binaries compared with the expected power from the as yet "unknown" population producing the CIB and CXB joint fluctuations as derived from their coherence by \citet{k19}. The grey, yellow and navy lines represent {\bf Early, Medium and Late} enrichment, respectively. The density of the lines once again represents the probability of realization of our model based on the sampling of the parameter space.}
\end{figure*}

The predicted power spectrum of the unresolved [2-5]$\mu$m CIB fluctuations from our model are shown in Fig.~\ref{fig:CIBPS} and that of X-ray sources in [0.5-2] keV range modeled here as accreting PBHs + X-ray Binaries, are shown in Fig.~\ref{fig:XPS}. These are compared with the expected power from the hitherto "unknown" population producing the CIB and CXB joint fluctuations as derived from their coherence by \citet{k19}. Once again, we show the realizations of the Early, Medium and Late enrichment models for IMF-3A, while the case for models IMF-3B and IMF-3C are reported in the Appendix. We have focused on the IMF-3A model and chosen to elaborate on it as the observed excess can be accounted for only in realizations of IMF-3A, regardless of the enrichment history. We point out that for IMF-3C, the  power spectrum predictions in the late enrichment case produce a sizeable fraction of realizations where the shot noise component exceeds the measure so we can safely conclude that IMF-3C provides the least likely scenario.

\begin{figure*}[!t]
\center
\includegraphics[width=0.7\textwidth]{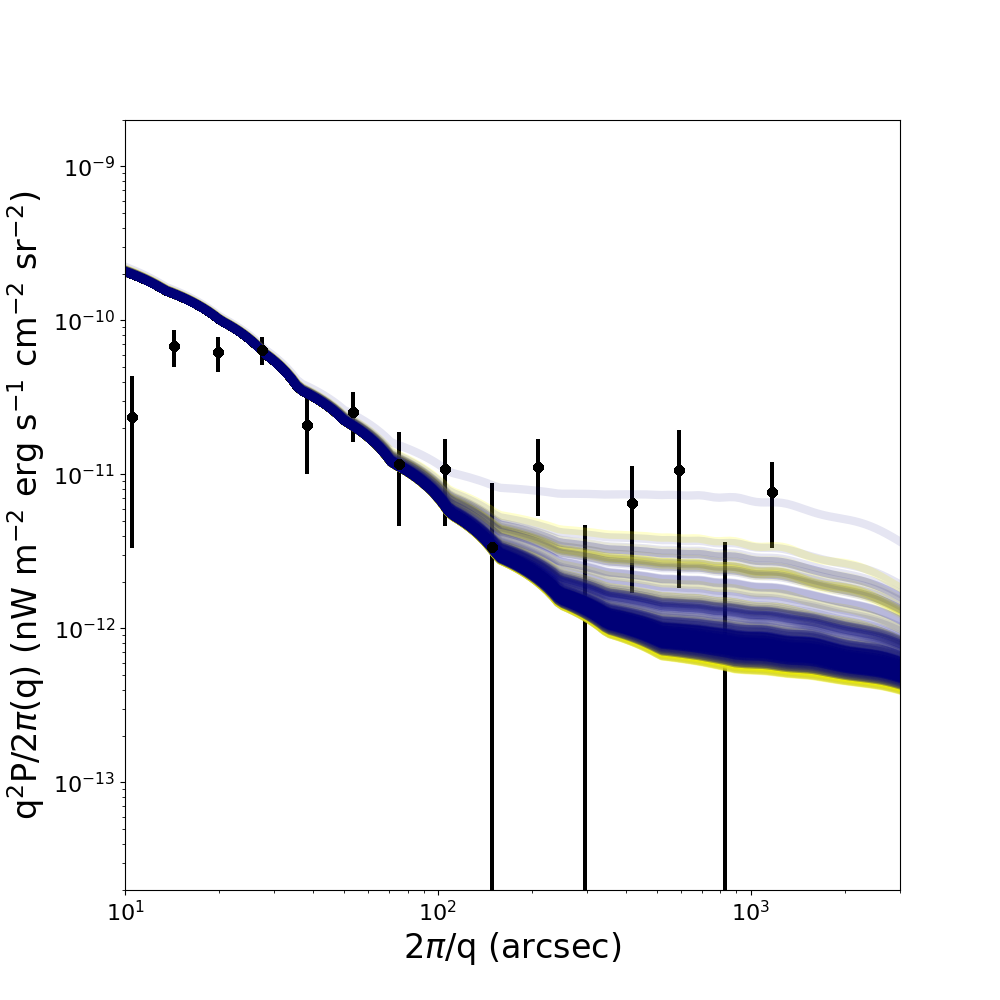}
\caption{\label{fig:IXPS} \label{fig:CIRPS} The cross-power spectrum of the unresolved [2-5] $\mu$m  CIB and [0.5-2] keV CXB fluctuations: computation from realizations of the IMF-3A model are added to the foreground galaxy estimate from \citet{helgason12}. Data points are a combination of the results of \citet{cap13,cap17}. As before, the grey, yellow and navy lines represent {\bf Early, Medium and Late} enrichment, respectively. The density of the lines represents the probability of realization of our model based on the sampling of the parameter space. Interestingly, several realizations from the IMF-3A model can successfully explain the excess observed over the foreground signal.}
\end{figure*}

The cross-power spectrum of the unresolved [2-5] $\mu$m CIB versus [0.5-2] keV CXB fluctuations from our model added to the foreground galaxy estimate from \citet{helgason12} is shown in Fig.~\ref{fig:IXPS}. The data points over-plotted are combined results from \citet{cap13,cap17}. We plot the three enrichment models and the three IMFs as before. Only a few realizations from IMF-3A and a few from IMF-3B with the Late/Medium enrichment model can explain the detected excess power on large angular scales in CIB versus CXB cross power spectrum. As a side note, our modeling shows that without the AGN component introduced into our model, the X-ray auto power and the cross-power would have been significantly weaker than what shown in Fig.~\ref{fig:IXPS} and Fig.~\ref{fig:XPS}. Therefore, our model strongly suggests the existence of such a population of sources rising at z$<$15.

\section{Discussion}
\begin{figure*}[!t]
\includegraphics[width=\textwidth]{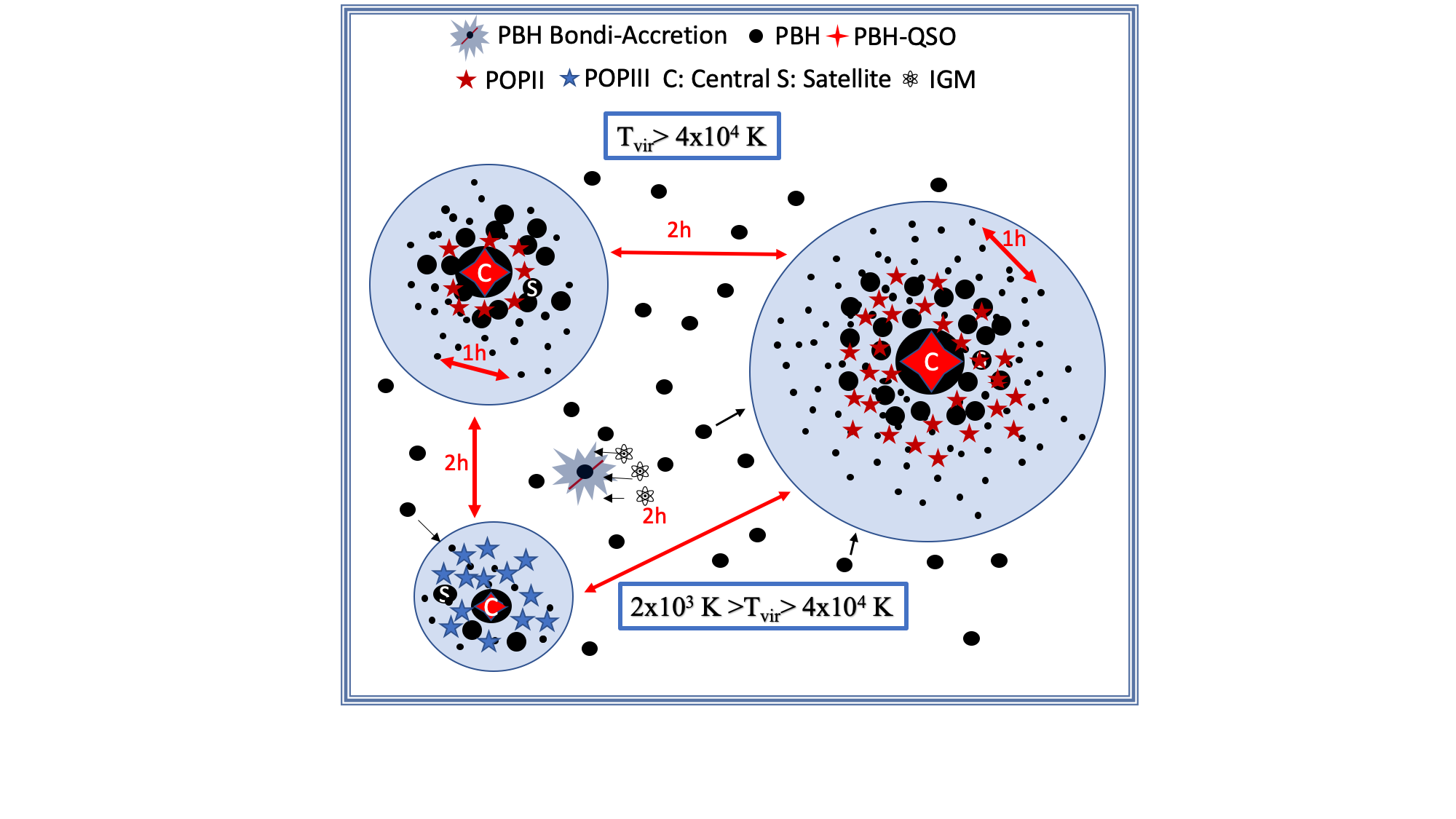}
\caption{\label{fig:schema} A schematic of our model of the early Universe in the PBH-\LambdaCDM~ Cosmology at z$\sim$10--15. PBH (black dots) initially stream with a low relative velocity with respect to the IGM allowing them to capture baryons while accreting via the Bondi mechanism, hence producing an early, high-z component of the CXB and a very faint CIB. Over time, PBHs clump to produce halos around the most massive PBHs. When the halo reaches the threshold circular velocity to capture baryons and commence cooling, stars begin to form. At the same time the high relative velocity of DM (PBHs) and baryons turns off Bondi accretion. The growth of structure at high-z in the PBH-\LambdaCDM~ Universe strongly favors early collapse of the lowest mass halos with T$_{vir}<$4$\times$10$^4$ K where gas is cooled by H$_2$, leading to the formation of massive Pop III stars. The higher mass halos cooling through atomic processes or halos polluted by metals produce Pop II stars. X-ray binaries inject more X-rays into the IGM. At the time when star formation commences, fresh baryons become available to power obscured AGN activity from gas accretion onto the most massive, central PBH-SMBH while lower mass satellite PBHs in the halo continue their Bondi accretion. These halos, whose mass is by definition proportional to the mass of the central BH, continue to grow while the SMBH grows and merges with other PBHs. In the cartoon shown here, we also show how we split the different components of the emission for purposes of calculating the clustering and the total emission. The central SMBH counts as the central source in the IR and X-ray emission, while smaller PBHs and stars fill the halo as satellites.}
\end{figure*}
In this paper we have explored the high-redshift properties of a Universe in which DM matter is composed of a broad mass spectrum of PBHs. The main picture that we postulate in this paper is summarized in Fig.~\ref{fig:schema}

The key results of our investigation of the PBH-\LambdaCDM~model can be summarized as follows: 

\begin{itemize} 

\item At z$>$6 low mass halos are more abundant than in the classical \LambdaCDM~ model even though a large fraction of these halos do not satisfy the physical conditions required for commencing star formation. The relative velocity of the IGM and DM suppresses star formation at M$_h<$ 10$^{5-6}$ M$_{\odot}$. However, depending on the enrichment history, our model predicts a secondary peak of star formation at z$\sim$15-20 beyond the well established observed peak at low redshifts, at z$\sim$3. This second peak at very early epochs is found to be driven by mini halos that most likely host the first episode of Pop III star formation. In terms of the overall star formation history and its evolution during the period that we have observational constraints, our model successfully reproduces the SFRD observed in deep surveys by \citet[e.g.][]{Bouwens} and modeling by \citet{madau}. This means that in terms of SFRD at z$<$6--7 our model predicts a scenario which is basically indistinguishable from predictions and measurement of \LambdaCDM. 

\item In order to explain the observed fluctuations in the CIB and CXB as a corollary of the PBH-\LambdaCDM~ scenario, we assumed that SMBHs are simply the most massive objects in each halo and by construction their mass scales as the mass of the host halo across cosmic time. These SMBHs then grow as obscured AGN with a product of their Eddington rate times the halo occupation fraction of a few percent. These assumptions lead to significant boosting of the X-ray fluctuations and hence of the unresolved CIB vs CXB  cross-power spectrum with a modest effect on the CIB fluctuations alone. In terms of accounting for the power in fluctuations, accreting AGN turn out to be lead players, as they inhabit higher mass halos which are extremely rare and  biased and, are at the same time, extremely luminous. The combination of these two properties makes AGN the objects of interest in our quest for the missing CXB power. 

\item In the PBH-\LambdaCDM~Cosmology we observe early production of the X-ray background due to the early commencement of Bondi accretion from the streaming IGM onto PBHs. This emission accounts for about 1\% of the unresolved CXB. A negligible fraction of the CXB is produced by XRBs during the star formation bursts associated with early star formation peaks. At z$<$15 the unresolved CXB becomes dominated by the onset of AGN activity that is a natural consequence of the PBH nature of DM. 

\item A hard constraint that any model of the early Universe has to comply with is the Thompson Optical depth that must not exceed measurements from current CMB experiments. Our model doesn't follow any detailed prescription for the reionizaton history, and we simply use \citet{planck} observations directly to drive the fit. While our model in general satisfies the observational limits, it produces several possible reionization histories depending on the combination of SFRD and f$_{esc}$ with high early star formation efficiency and low f$_{esc}$ less favored. In fact a high star formation efficiency paired with extremely low f$_{esc}$ in Pop III halos would trigger extremely early reionization. Further observations, in particular at 21 cm are a required addition to the model to predict detailed ionization histories to constrain the model. 

\item A polarization signal is predicted to be produced in the CMB during reionizaton. The large-scale E-mode CMB polarization can therefore serve as a very sensitive probe of the reionizaton history \citep{2016ASSL..423..227R}. If reionizaton starts significantly earlier than is classically assumed via instantaneous recombination around $z=6-7$, there will be a detectable peak in the EE component of the polarization at higher multipoles than predicted in the case of \LambdaCDM~ \citep[see e.g. Fig.~8 of][]{k19}. Similarly, the additional radio background produced by accreting PBHs at higher-z can imprint a signal in the spin temperature of neutral hydrogen that could be discerned in the integrated, redshifted 21-cm signal from high-precision low frequency radio wave measurements extending the EDGES experiment \citep{2018Natur.555...67B}.

\item A common feature of our predictions across all models is a steepening of the NIR source counts at m$_{AB}\,>\,28-30$. Our model clearly shows that if the IMF of Pop III stars is tilted to be top-heavy, the steepening occurs at brighter magnitudes than in the case of classic Pop III Kroupa IMF or the Lognormal Pop III IMF. This steepening appears at the 10 ks exposure limit for JWST. However, deeper observations and exploiting the lensing magnification of sources behind clusters will be extremely powerful in detecting these fainter counts. 

\item The PBH-DM with the wide mass distribution assumed here, will cluster around supermassive PBHs in the centers of DM haloes. Similar to the dynamical hardening processes in stellar globular clusters, many of these PBHs will merge through triple interactions, while others will be ejected. In particular, mergers between intermediate mass black holes and SMBHs, as well as extreme mass ratio inspirals will be more frequent and will occur at earlier redshifts than expected for classical \LambdaCDM. The gravitational wave observatories LISA and PTA will be sensitive to the signal from these expected high redshifts and can also provide strong constraints on the PBH-DM scenario presented here. 

\item As demonstrated, the excess CIB fluctuations can be explained by a stellar origin  in some realizations of IMF-3A, and some additional amplitude can be recovered by adding an accreting AGN component. However, while we showed that by adding an AGN component we satisfy the requirements posed by the CIB vs CXB cross-power, full accounting for the required integrated 1 nW m$^{-2}$ sr$^{-1}$ measured signal \citep[see e.g.][]{helgason16,k19} is not achievable with a stellar component alone. Some of this can be mitigated by including modeling potential additional unaccounted components like a more detailed measurement of the cirrus and/or a IHL component. In the case of the PBH-\LambdaCDM~ scenario explored here, where the typical mass of the PBHs is assumed to be of the order 1 M$_{\odot}$, the excess number of mini-halos is insufficient to produce enough stars to boost the CIB fluctuations on large angular scales. However, other scenarios with larger average PBH masses that have not been explored here, might produce more star forming halos at even higher masses at lower redshift (and hence with larger bias) and therefore impact the fluctuations significantly. Such high masses for PBHs for instance, can be obtained by assuming that PBH-PBH mergers shift the mass distribution shown in Fig.~\ref{fig:fPBH} to higher masses over time and therefore lead to the collapse of more massive halos - a modifying effect that we plan to pursue in future work. Finally, it bears noting that because of the high shot noise predictions, the majority of IMF-3C realizations are ruled out. \\

\item It turns out that the X-ray spectral energy distribution of the CXB/CIB cross-correlation signal also contains information about their production mechanism, for example, on the amount of absorbing material around the accreting black holes. \citet{li18} show a model comparison for the averaged signal from all Chandra deep fields, emphasizing that in particular the signal in the softest X--ray band (0.5--1 keV) can provide useful discrimination of the sources of production. However, the statistical quality of these data are insufficient to do so at the present time. Future observations, in particular the cross-correlation of the data from eROSITA deep fields -- and those from the planned ATHENA satellite, along with Euclid NIR observations, will in combination provide powerful constraints and tests of our model \citep{2019ApJ...871L...6K}.

\item We note that the models explored here look like DCBH models with early BH formation as AGN turn on very early in this scenario too. The PBH-DM models predict an initial occupation fraction of unity - one BH per galaxy naturally. This of course, stands to be modified at later times due to dynamical interactions arising from binary black hole mergers. The DCBH formation channel stands to be tested with data from the NIRCAM and MIRI detectors on JWST, that will soon be available \cite{Natarajan+2017}. 

\item In terms of further AGN predictions of the PBH-\LambdaCDM~ model, we note that extrinsic AGN variability on multiple time-scales due to ubiquitous microlensing is expected to be more prevalent than in \LambdaCDM. If and when high-z microlensing signatures are detected, further support for the PBH-\LambdaCDM~ model is likely. Though a calculation is hard to set-up, given the granularity of PBHs in every DM halo in this model, we can expect many more gravitational wave coalescence events arising from 3-body interactions \citep[see e.g.][]{gow,gb17}.\\
\end{itemize}

To conclude, in this paper we show that a broad mass spectrum for PBH-DM can be easily accommodated in a model of the early Universe, producing features comparable to \LambdaCDM. One of the main limitations of this is work is the widespread lack of observational data to tightly constrain our parameter space at early epochs, times during which this model maximally differs from \LambdaCDM. In particular, the reionizaton history of the Universe is largely unknown and future instruments like JWST and SKA will likely provide a clearer window into this epoch. At the same time, CIB and CXB fluctuation measurements are currently limited to angular scales of a few tens of arcminutes, making them very susceptible to low counts statistics and sample variance. Future wide field surveys like those planned with the EUCLID, WFIRST-Roman, eROSITA and Athena missions, will finally permit the measurement of the fluctuations on several degree scale with high accuracy of a few percent. The lack of knowledge of the very high-z SFRD is likely filled in soon by the forthcoming launch of JWST.  Deep JWST data will provide a brand new window and allow us to explore star formation and early growth of AGN up to z$\sim$15. The detection of these high-z sources stands to inform our understanding of not just $f_{\rm PBH}$ but also the efficiency of star formation and the escape fraction at these early epochs, as these are the directly observable signatures predicted by modeling work here.

Finally, as mentioned above, PBH mergers are expected to be very frequent in this scenario and forthcoming LIGO runs and planned LISA, PTA and 3rd generation ground-based gravitational wave detectors will help in further testing this alternative DM scenario on this front. While in some respects this DM model might appear more baroque than standard \LambdaCDM~ as it involves multiple-mass components - in contrast to the single mass particle CDM - the explanatory power of the scenario makes it compelling. For instance, this is the only model that successfully accounts for the natural formation of early BH seeds; the existence of a correlation between properties of the central SMBH and the host galaxy and underlying dark matter halo and explains the CXB-CIB cross-correlation and its excess, while being consistent with all the multi-wavelength observations at $z\,<\,6$. Other alternate DM models, like those that include dissipative DM with self-interactions that lead to core-collapse (SIDM variants) are amongst non-CDM proposals for BH seed formation \citep[see for instance, recent proposal by][]{zurek+2021}. In the absence of any detection of the putative DM particle after several decades of direct and indirect experimental searches, PBH-DM offers an economical scenario that is well motivated by physics and is one that couples early-Universe physics with phenomena on cosmological scales in the late Universe.

\acknowledgements
We thank the anonymous referee for the timely reading and the useful comments that really improved this paper. NC acknowledges support from the Chandra-SAO Grant TM9-20008X and kindly acknowledges the University of Miami College of Arts and Science for support. NC kindly acknowledges the Cantera, Moscetti, and Larrea family for support. NC thanks Kari Helgason for valuable help in developing the model and Fabio Vito for providing extrapolations of the luminosity function. NC and GH thank INAF-OAS Bologna for kind hospitality in Summer 2021 and for providing convenient office space during the preparation of this paper. The authors thank Sasha Kashlinsky for insightful discussions and for providing the CXB fluctuation estimates. NC and GH kindly acknowledge the LIBRAE team for fruitful discussions. We also thank Juan Garc{\'\i}a-Bellido and Fernando Atrio-Barandela for helpful discussions. PN acknowledges the Black Hole Initiative (BHI) at Harvard University, which is supported by grants from the Gordon and Betty Moore Foundation and the John Templeton Foundation, for hosting her. PN is grateful for the Zoom platform for enabling continued collaborative work with NC and GH during this past difficult pandemic year. GH thanks the city of Remscheid, Germany, for the award of the R\"ontgen Medal and their hospitality during the days when this manuscript was submitted.

\bibliographystyle{aasjournal}
\bibliography{sample63.bib}

\appendix
\counterwithin{figure}{section}

\section{Posterior distributions}

In this section we show the posterior distributions derived by our MCMC chains. In all figures we plot the confidence contours for the 16\%, 50\% and 84\% quartiles. Each panel represents each combination of the possible 9 IMF and enrichment models.

\begin{figure*}[!t]
\includegraphics[width=0.33\textwidth]{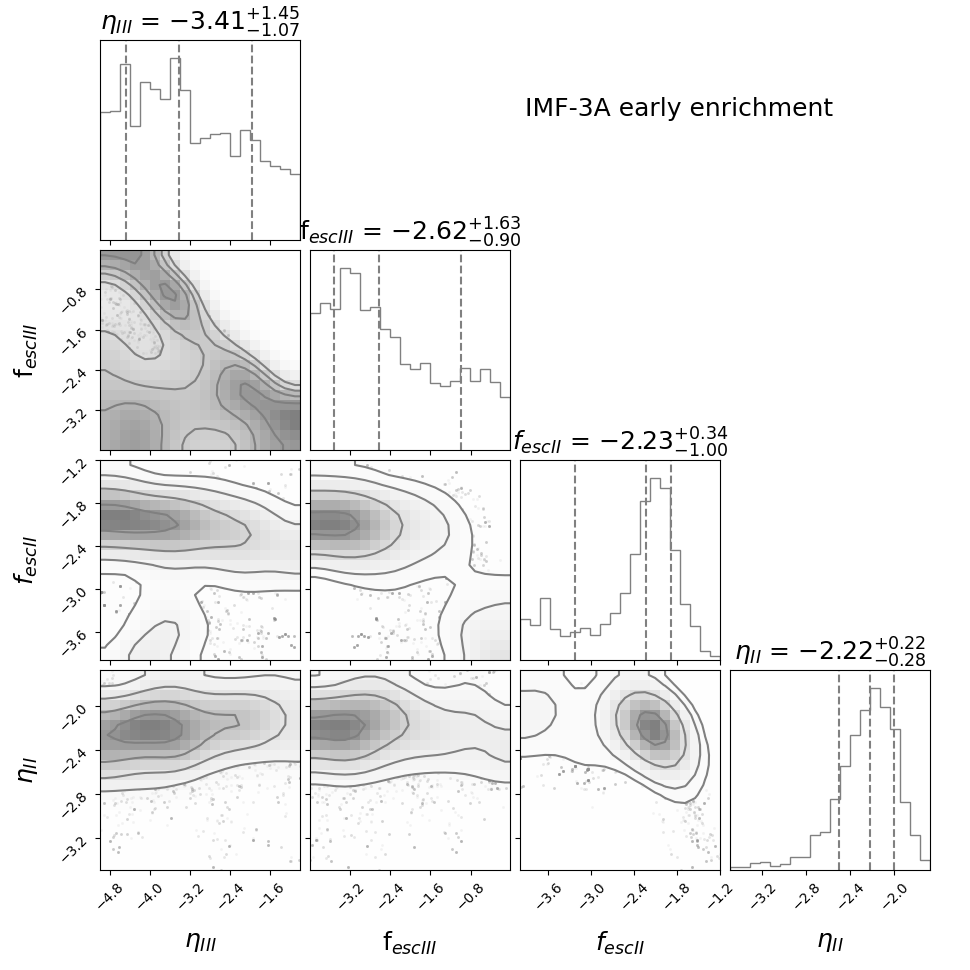}
\includegraphics[width=0.33\textwidth]{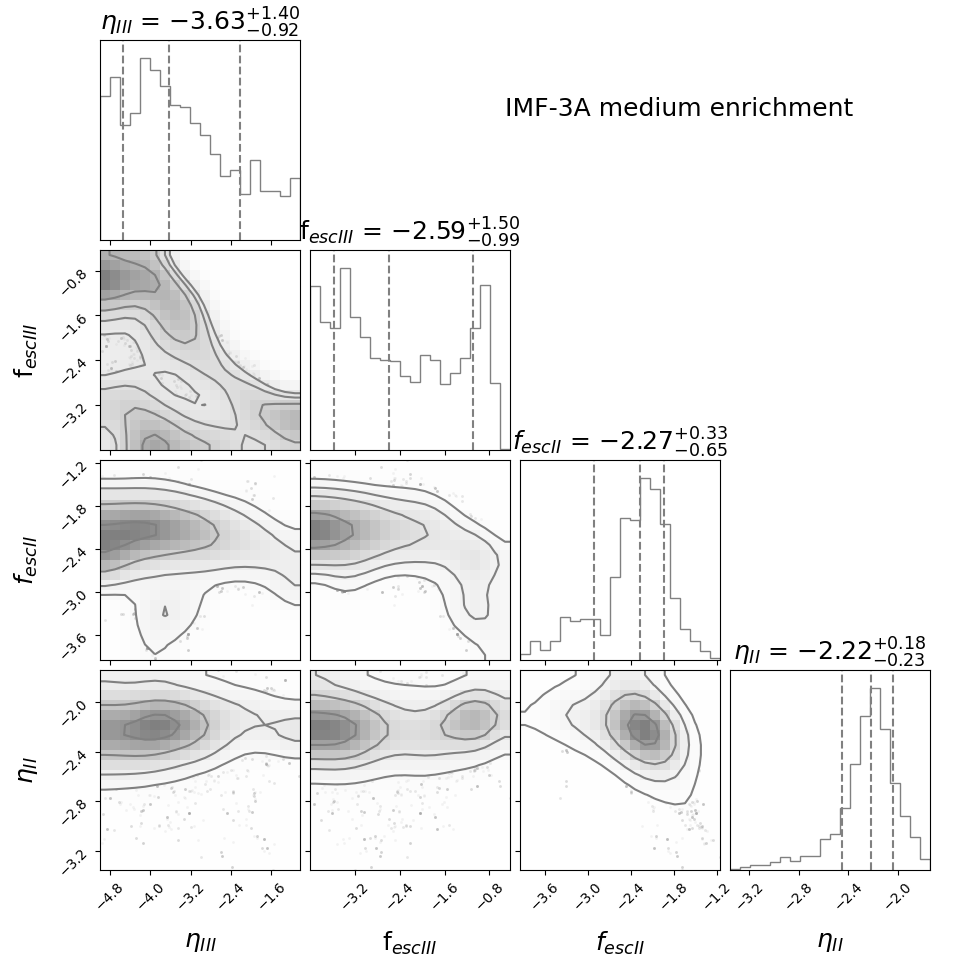}
\includegraphics[width=0.33\textwidth]{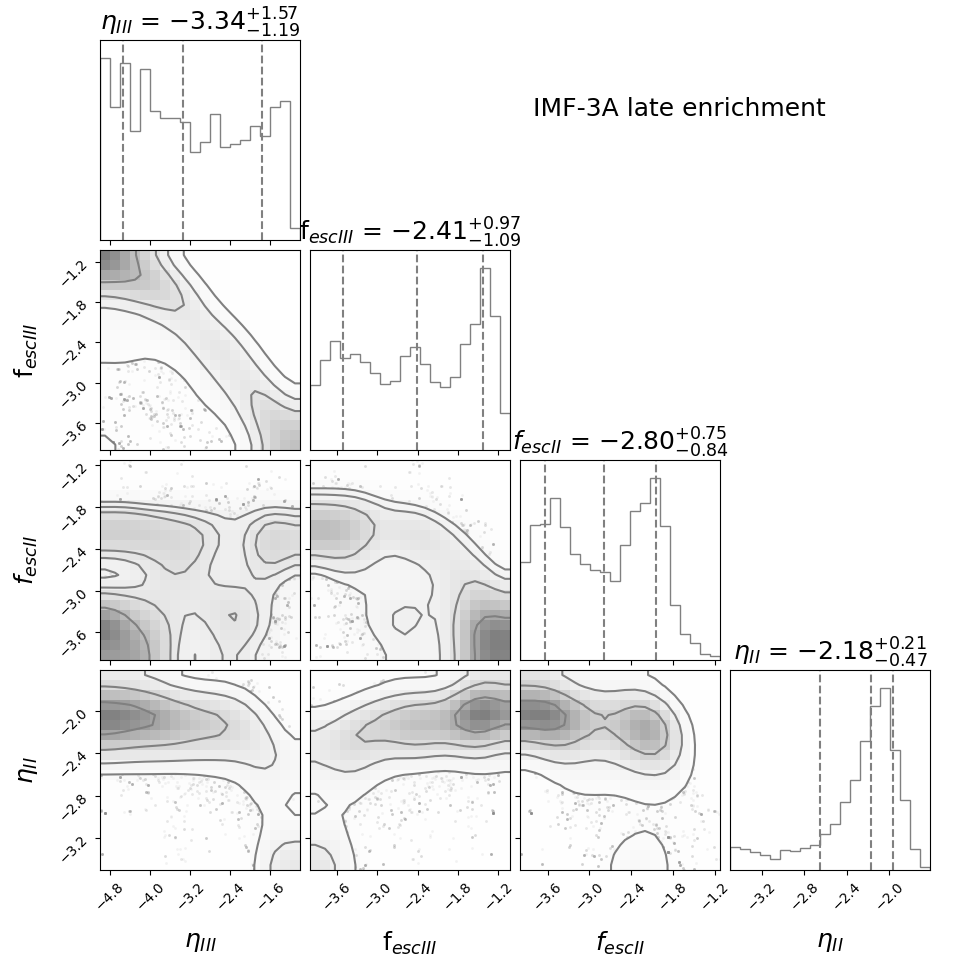}
\includegraphics[width=0.33\textwidth]{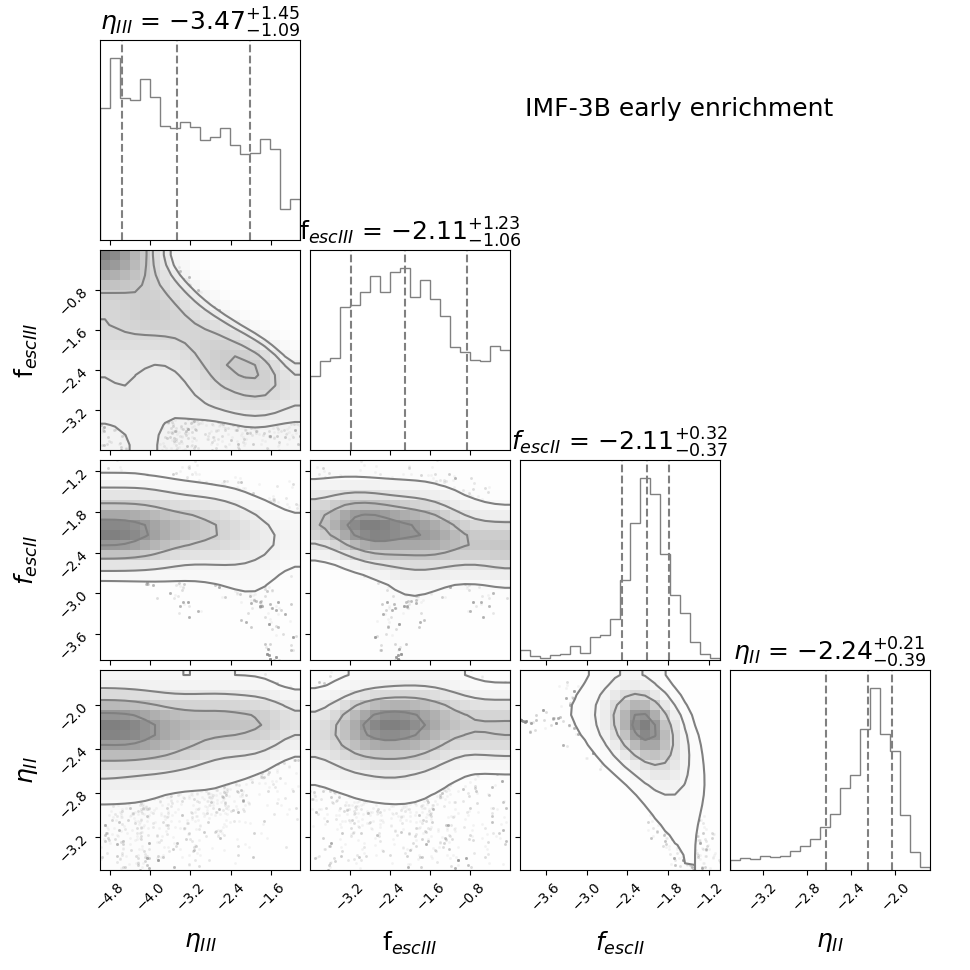}
\includegraphics[width=0.33\textwidth]{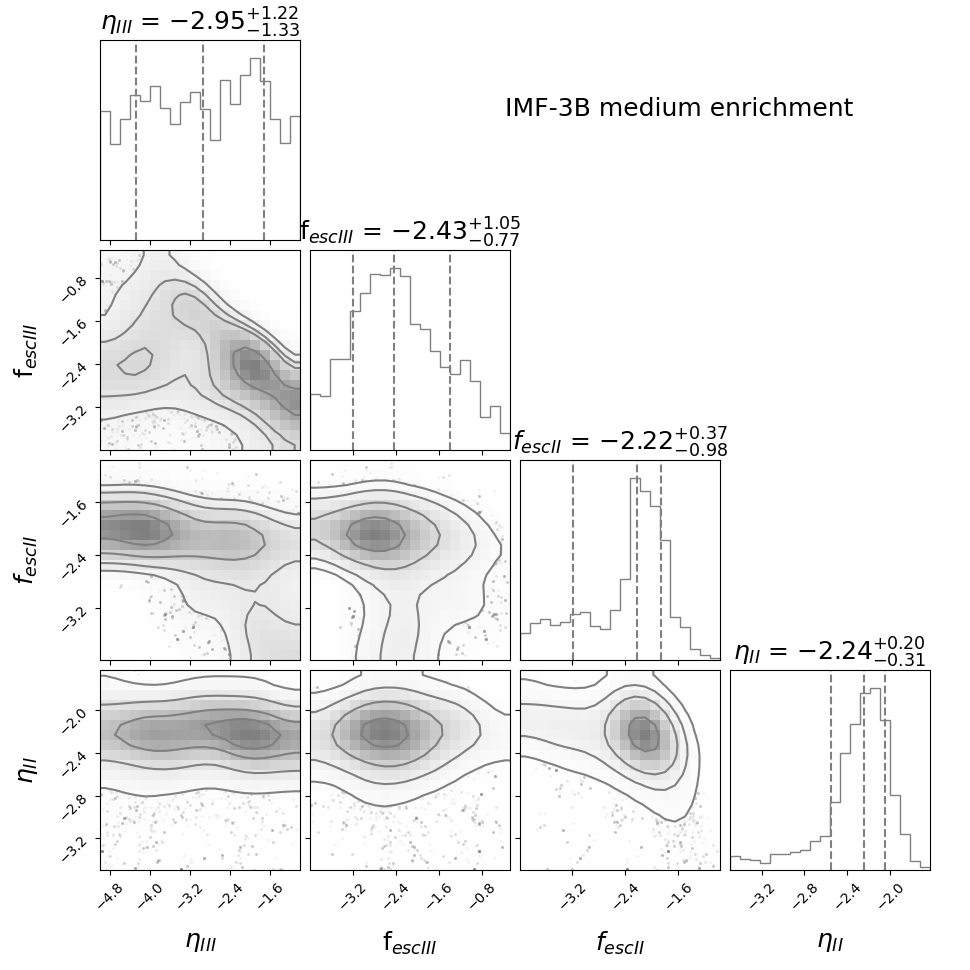}
\includegraphics[width=0.33\textwidth]{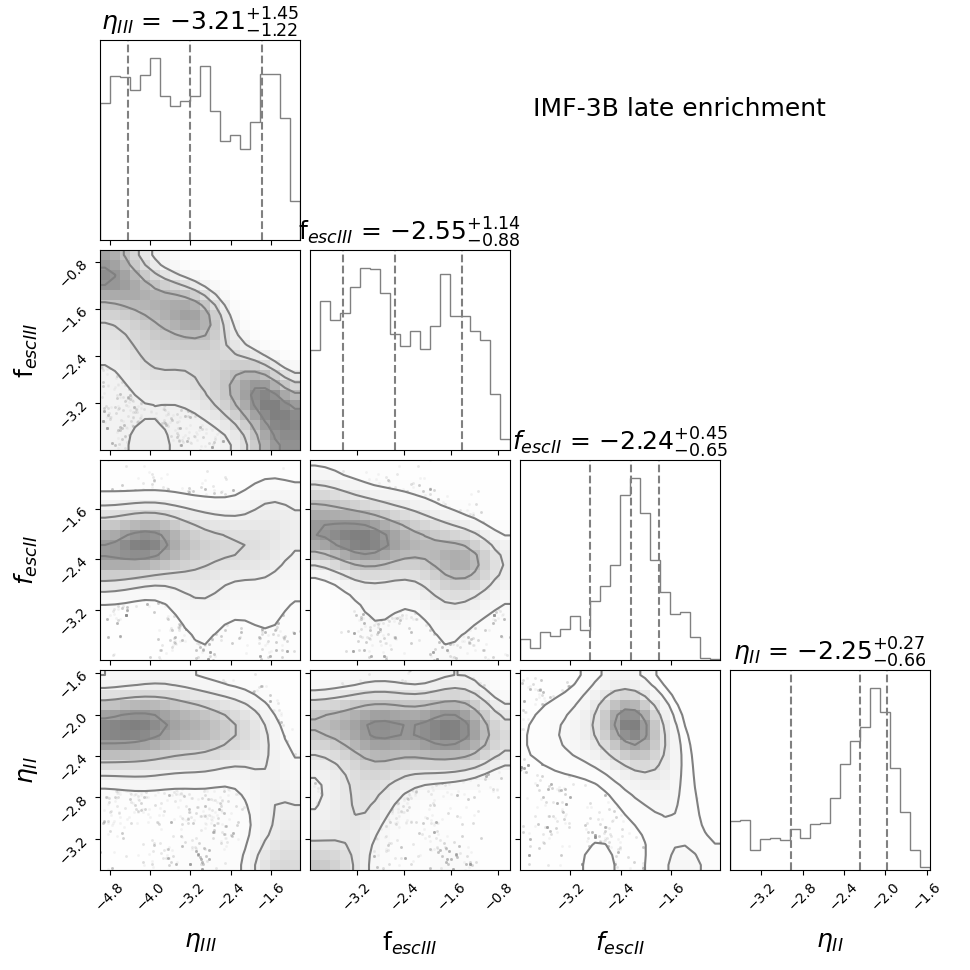}
\includegraphics[width=0.33\textwidth]{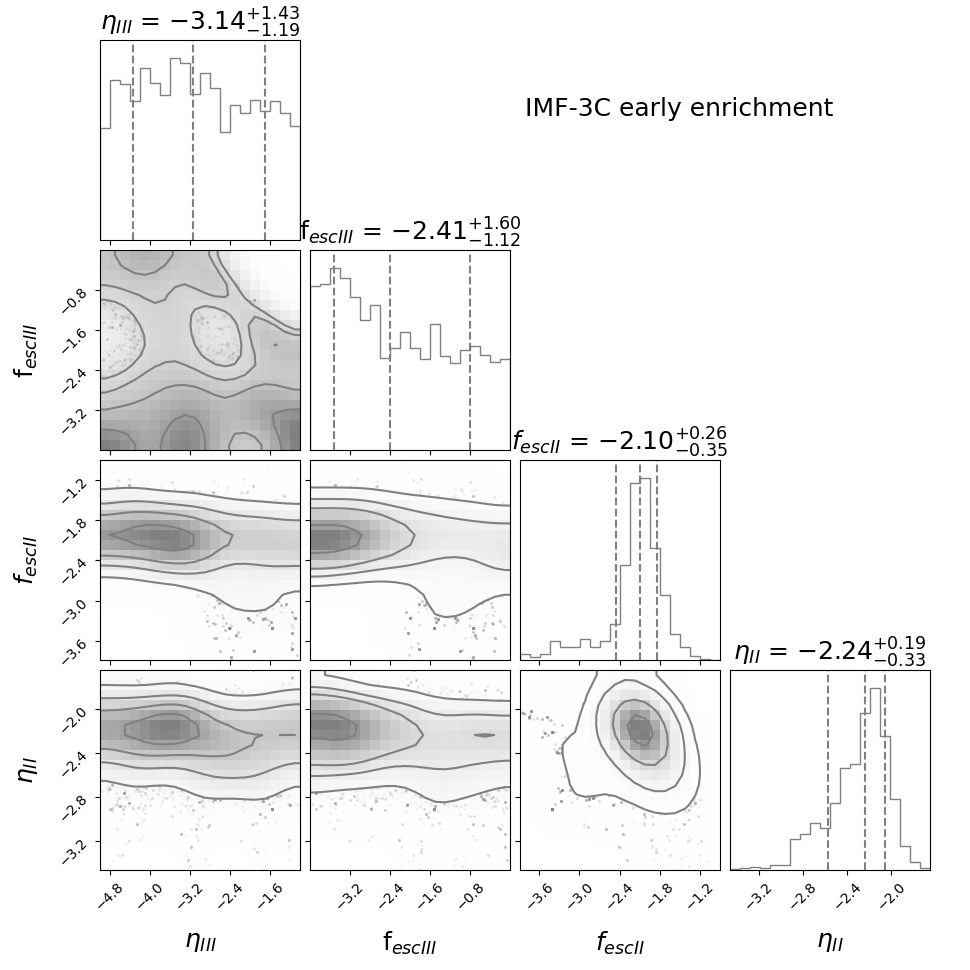}
\includegraphics[width=0.33\textwidth]{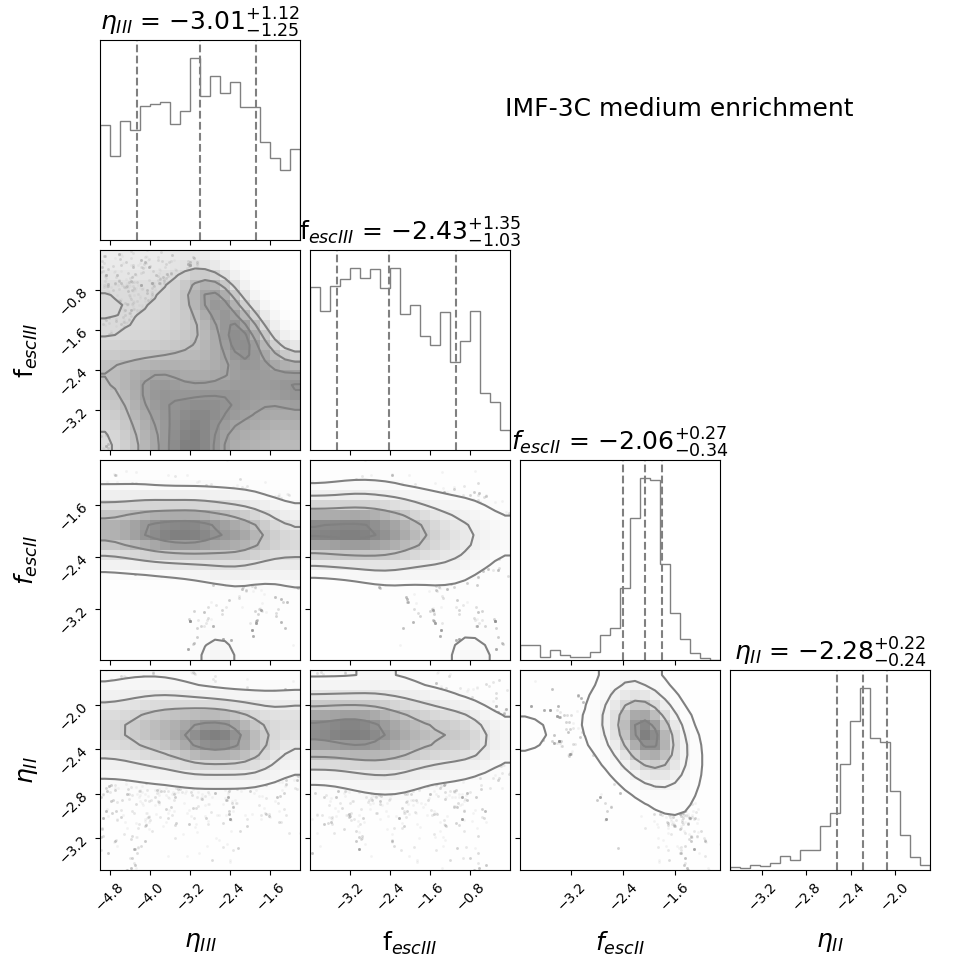}
\includegraphics[width=0.33\textwidth]{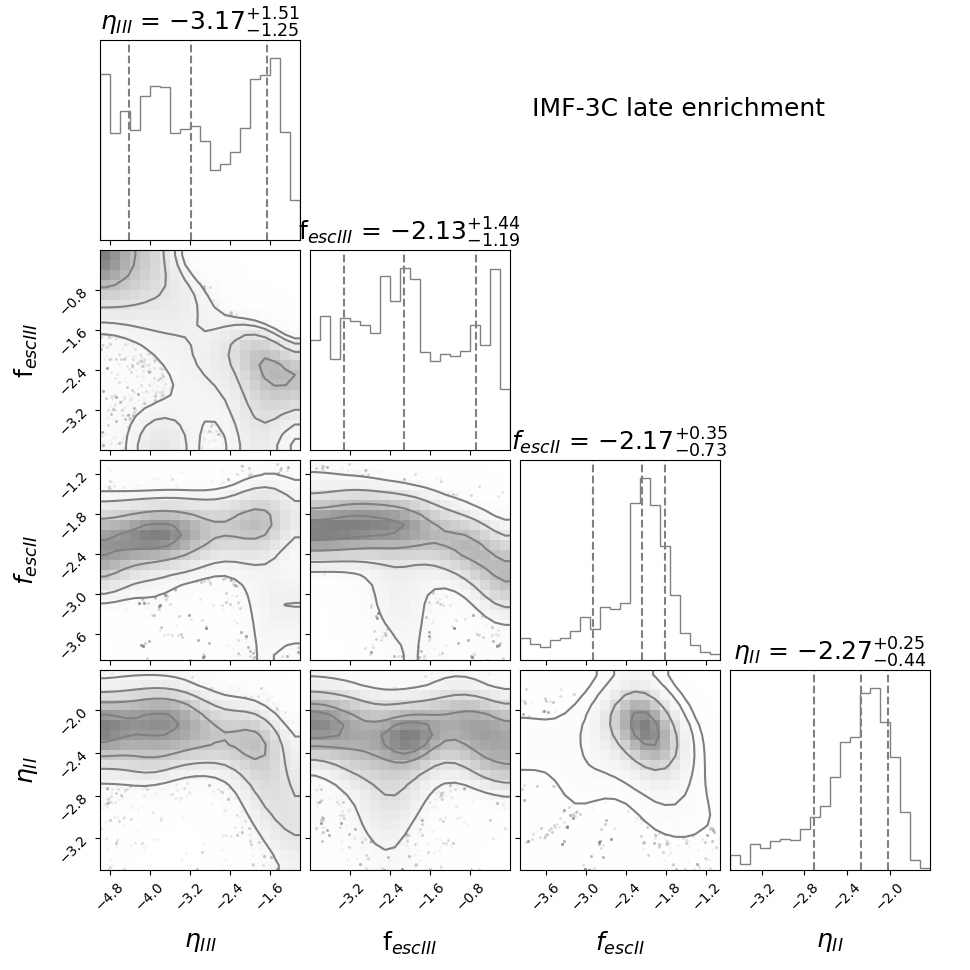}
\caption{\label{fig:app}  In shades of grey we plot the confidence contours for the Log$_{10}$ of  f$_{esc,II}$ , f$_{esc,III}$,  $\eta_{II}$, $\eta_{III}$. Rows, from top to bottom, represent the models IMF-3A, IMF-3B, and IMF-3C, respectively. Columns, from left to right, represent {\bf Early, Medium \& Late enrichment}, respectively.}
\end{figure*}

\section{Realizations for the IMF-3B and IMF-3C models}

In this section we provide the plots with results of the IMF-3B and IMF-3C model realizations. 

\begin{figure}[h]
\center
\includegraphics[width=0.45\textwidth]{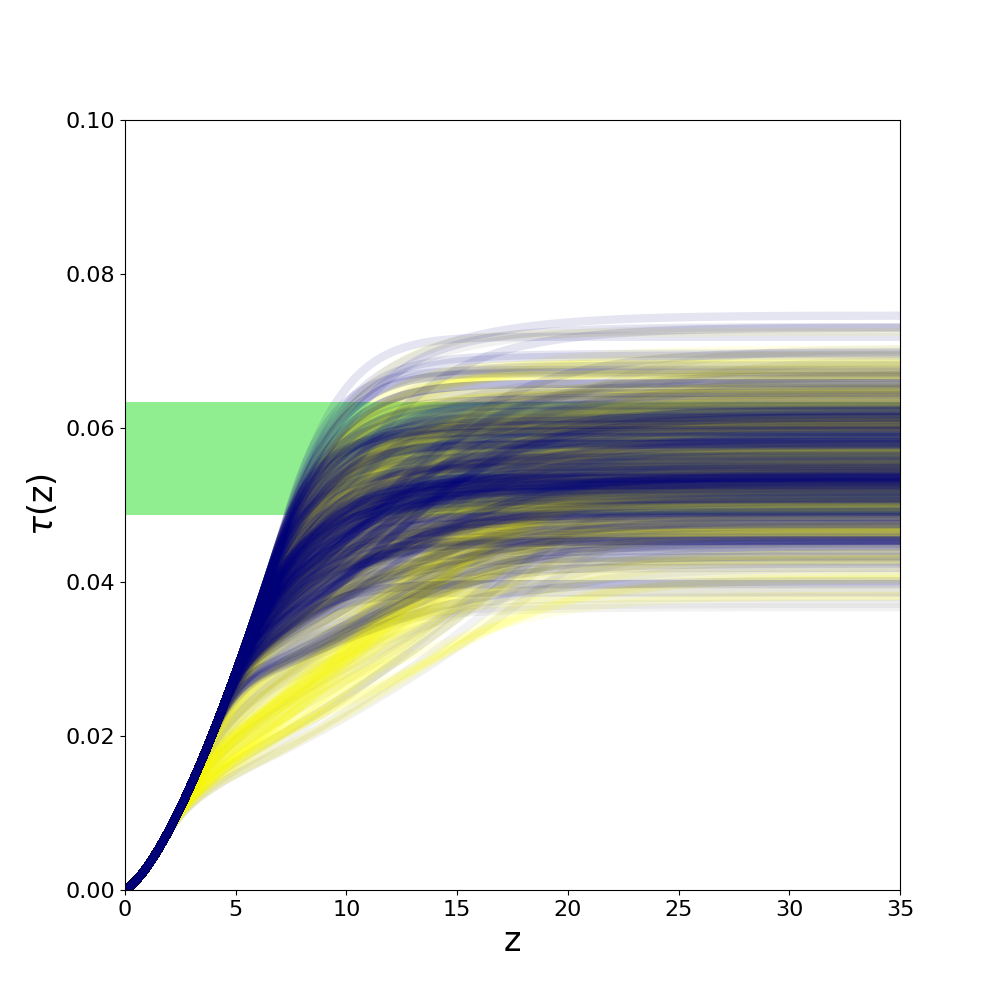}
\includegraphics[width=0.45\textwidth]{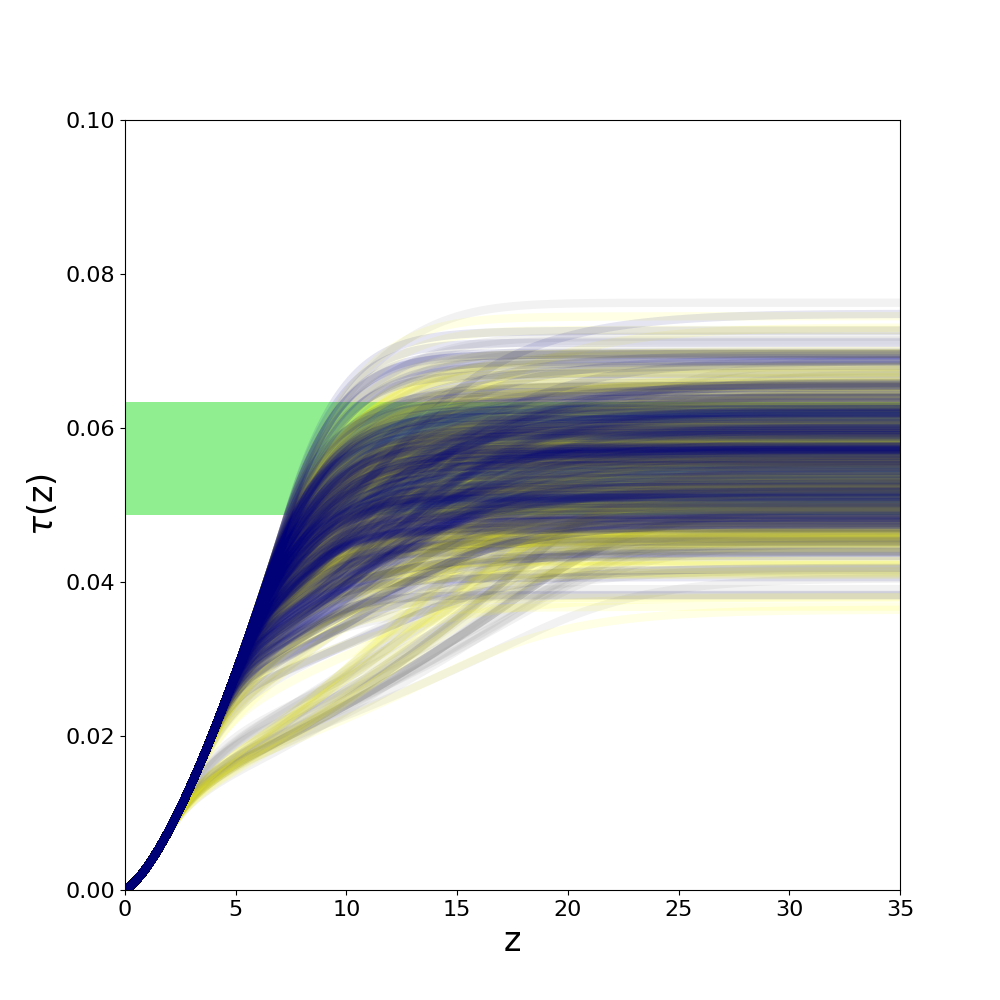}
\caption{The redshift evolution of the Thompson optical depth compared with Planck limits \citep{planck} represented as a green band. The two panels represent from $left$ to $right$ the IMF-3B and IMF-3C models. The grey, yellow and navy lines represent {\bf Early, Medium \& Late} enrichment, respectively. As in the text, the density of the lines represents the probability of realization of our model based on our sampling of the parameter space. Our model predicts many feasible reionizaton histories depending on the combination of f$_\star$ and f$_{\rm esc}$.}
\end{figure}
\begin{figure}[h]
\center
\includegraphics[width=0.45\textwidth]{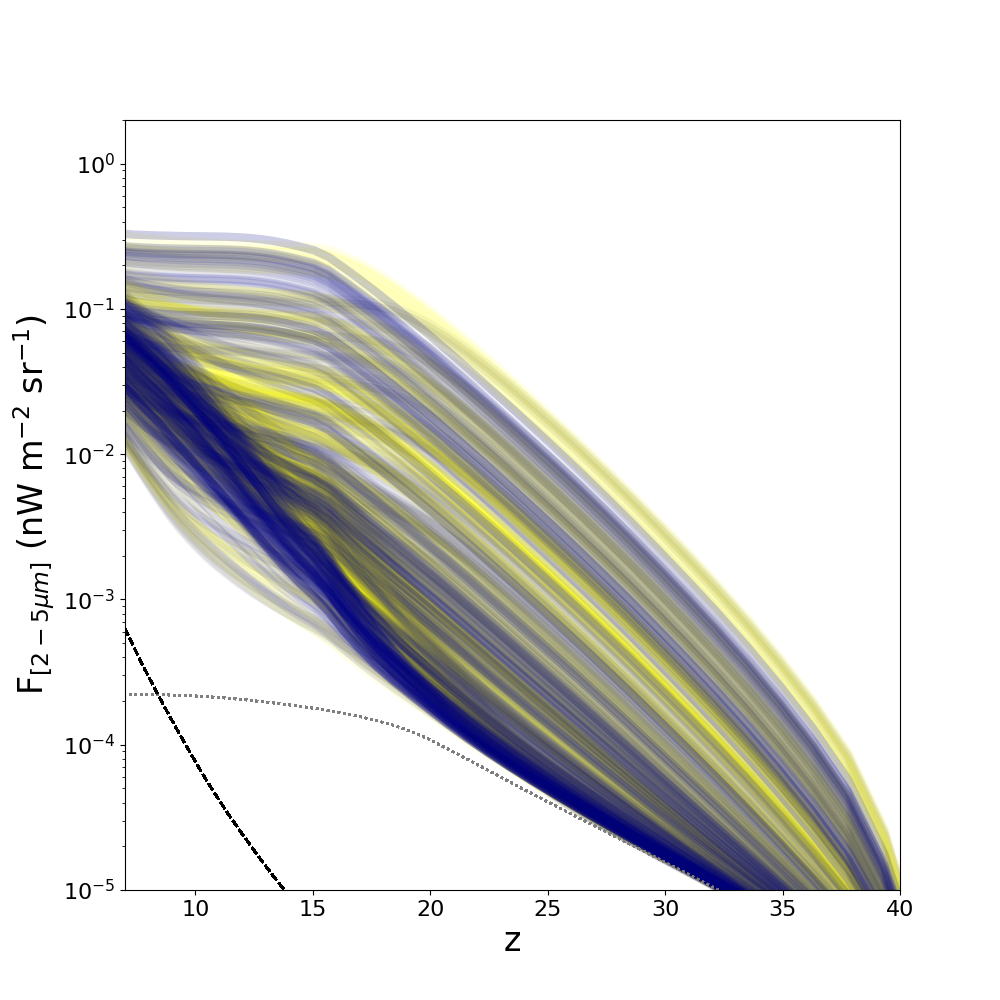}
\includegraphics[width=0.45\textwidth]{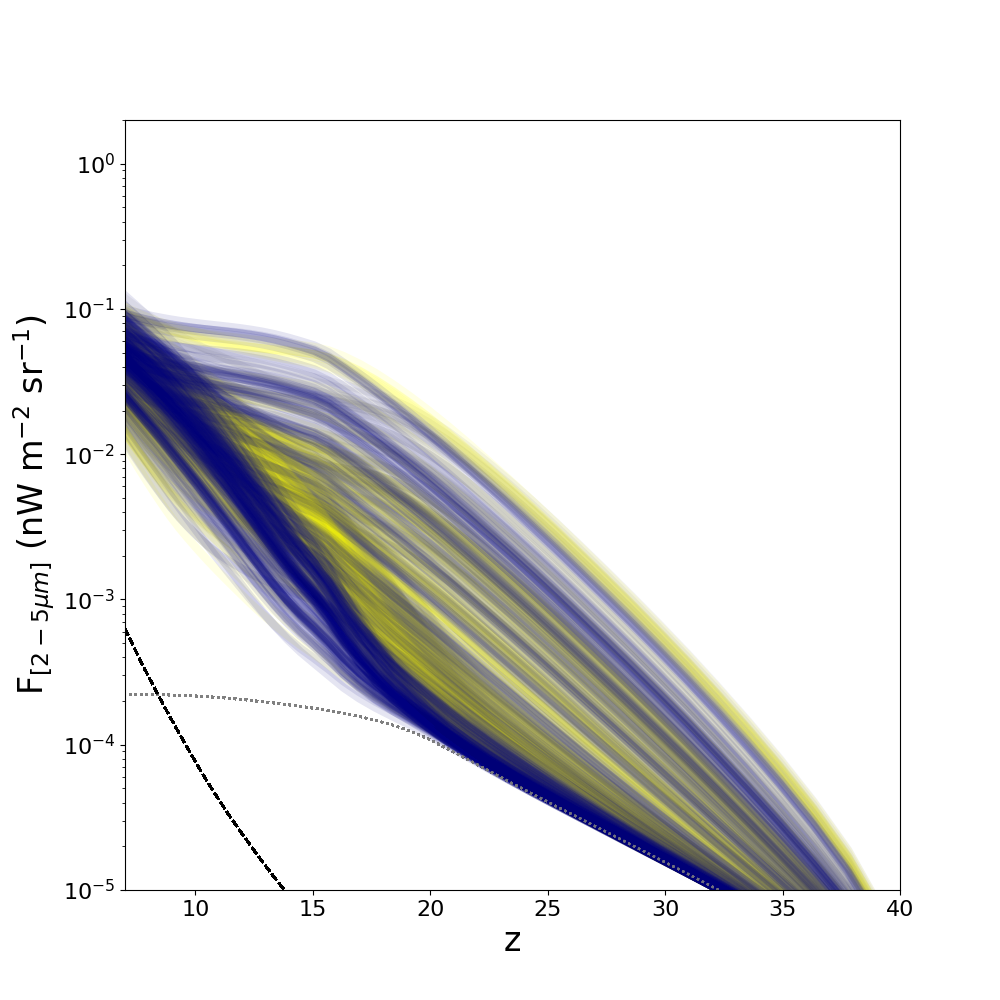}
\caption{The cumulative CIB flux production rate a function of redshift. The grey, yellow and navy lines represent {\bf Early, Medium \& Late} enrichment, respectively. The black dashed line represents radiation produced by AGN while the black dotted line represents accretion from streaming baryons onto PBHs. The two panels show from $left$ to $right$ the models IMF-3B and IMF-3C. As before, the density of the lines represent the probability of realization of our model based on the sampling of the parameter space. The bulk of the CIB is produced by star formation while AGN are subdominant.}
\end{figure}

\begin{figure}[h]
\center

\includegraphics[width=0.45\textwidth]{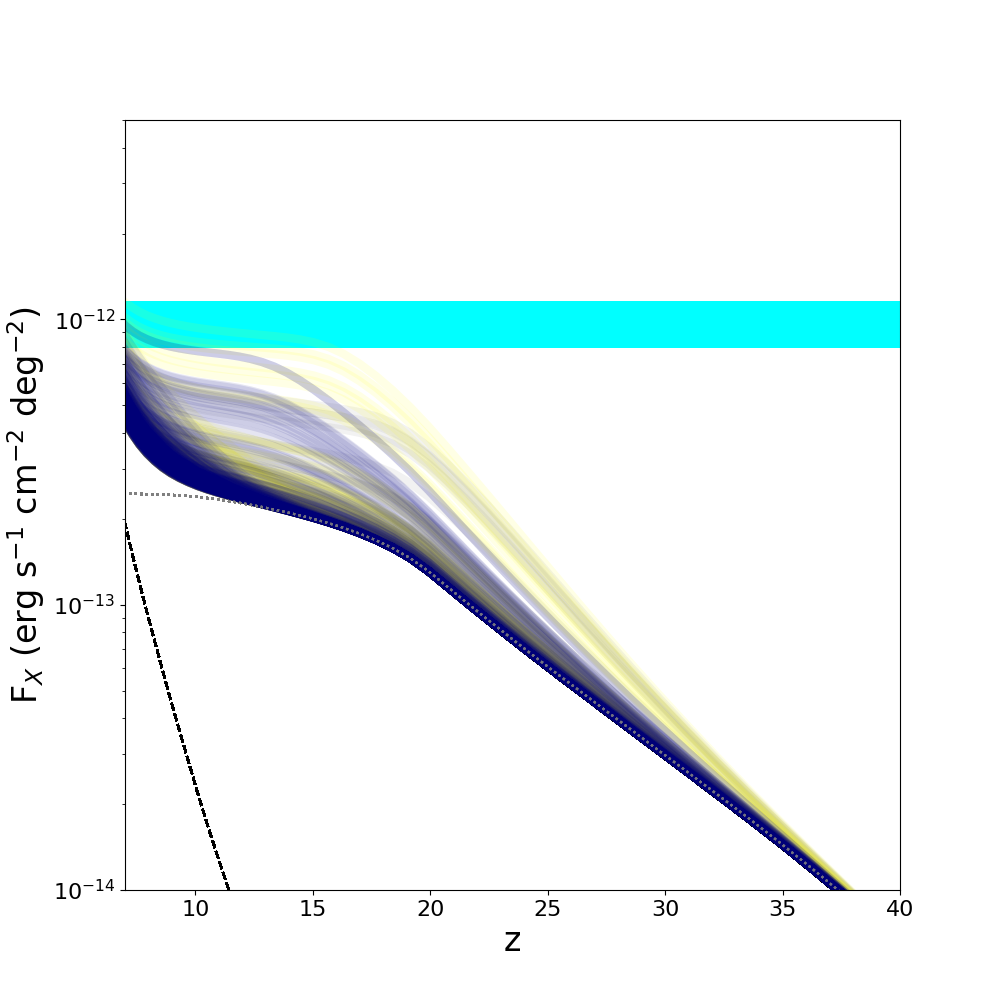}
\includegraphics[width=0.45\textwidth]{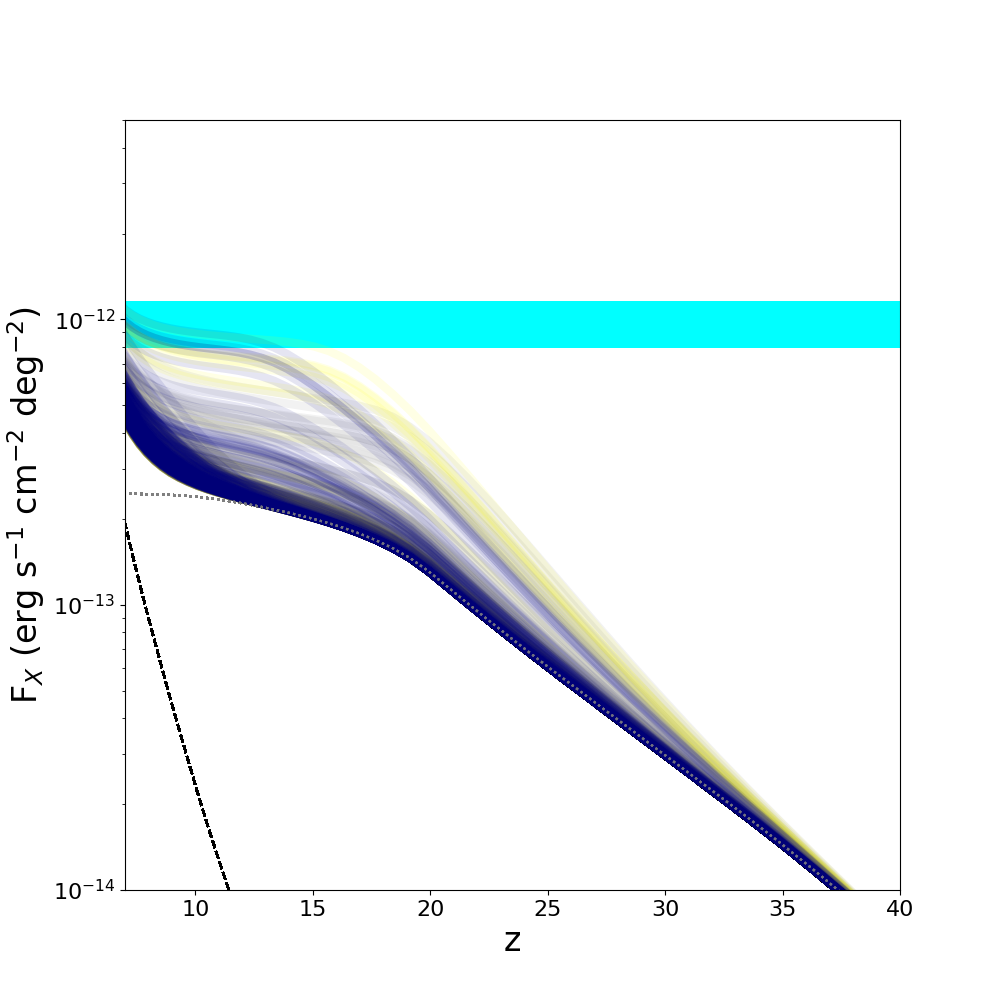}
\caption{The cumulative CXB flux production rate as a function of redshift. The grey, yellow and navy lines represent {\bf Early, Medium \& Late} enrichment, respectively. The black dashed line represents radiation produced by AGNs while the black dotted line represents accretion from streaming baryons onto PBHs.  The density of the plotted lines represents the probability of realization of our model based on the sampling of the parameter space. The $Cyan$ band represents the current limit on the unresolved CXB from \citet{cap17} and \citet{hm06}. The two panels from $left$ to $right$ represent the IMF-3B and IMF-3C models. At z$>$15 the bulk of the CXB is produced by accretion onto PBH satellites while, at z$<$15 central AGN become dominant. X-ray binaries meanwhile contribute only of the order of a few percent of the CXB.}

\end{figure}

\begin{figure}[h]
\center

\includegraphics[width=0.45\textwidth]{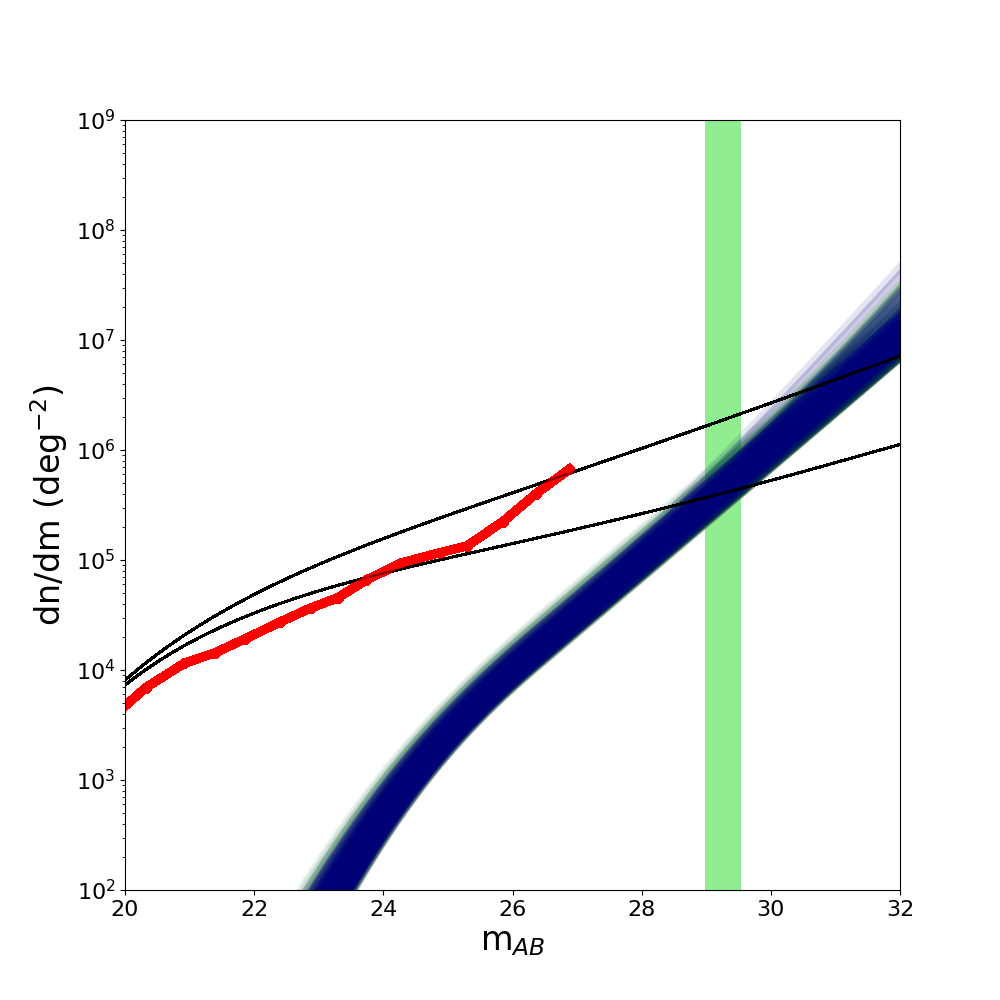}
\includegraphics[width=0.45\textwidth]{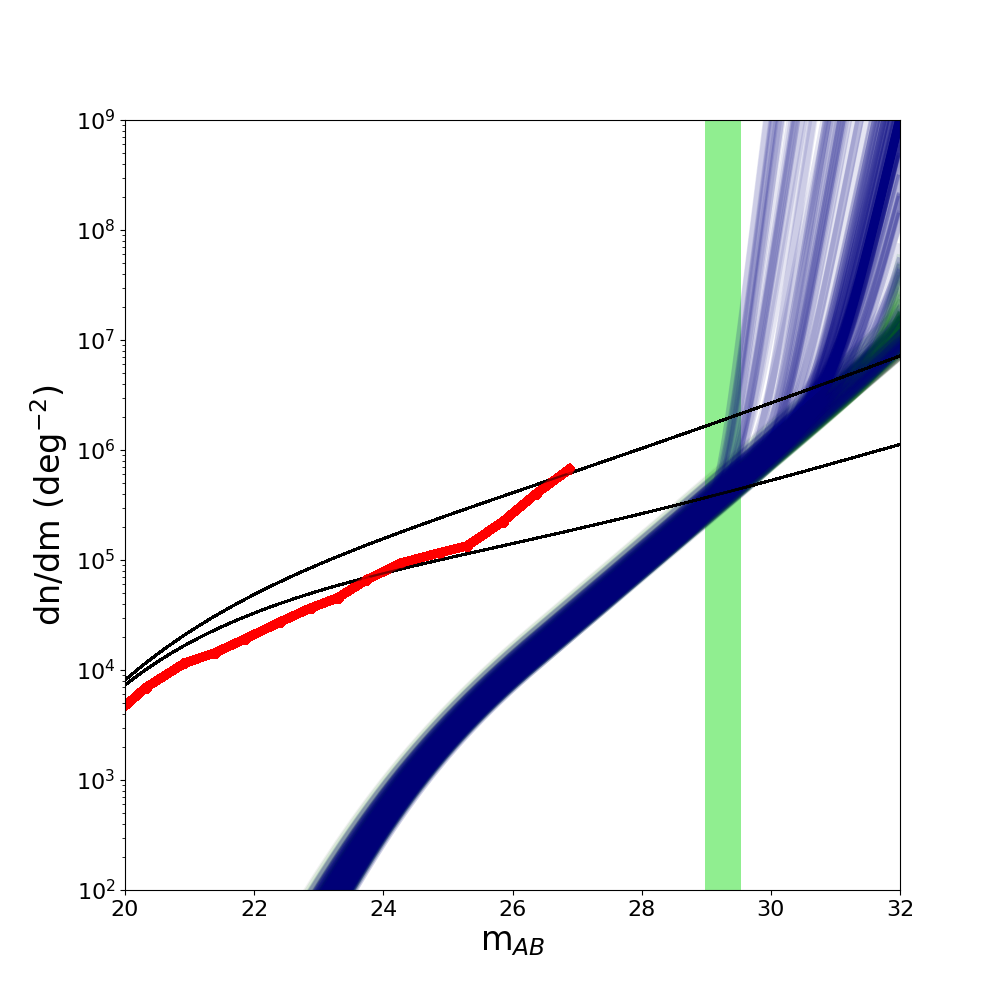}
\caption{The predicted source counts for $z\,>\,6$ NIR sources. The grey, yellow and navy lines represent {\bf Early, Medium \& Late} enrichment, respectively. The predictions are compared with S-CANDELS data by \citet[][]{2015ApJS..218...33A} (red-line) and with the extrapolations of the population synthesis model of \citet{helgason12} based on the faint end of the luminosity function. The two panels represent from $left$ to $right$ the IMF-3B and IMF-3C models. All the models predict comparable and compatible results with a steepening of the counts at $m_{\rm AB} > 28-30$ with a slightly earlier onset of this high-z population in the IMF-3A model. The green band represents the JWST 10 ks magnitude limit calculated for a 10ks exposure.}
\end{figure}

\begin{figure}[h]
\center

\includegraphics[width=0.45\textwidth]{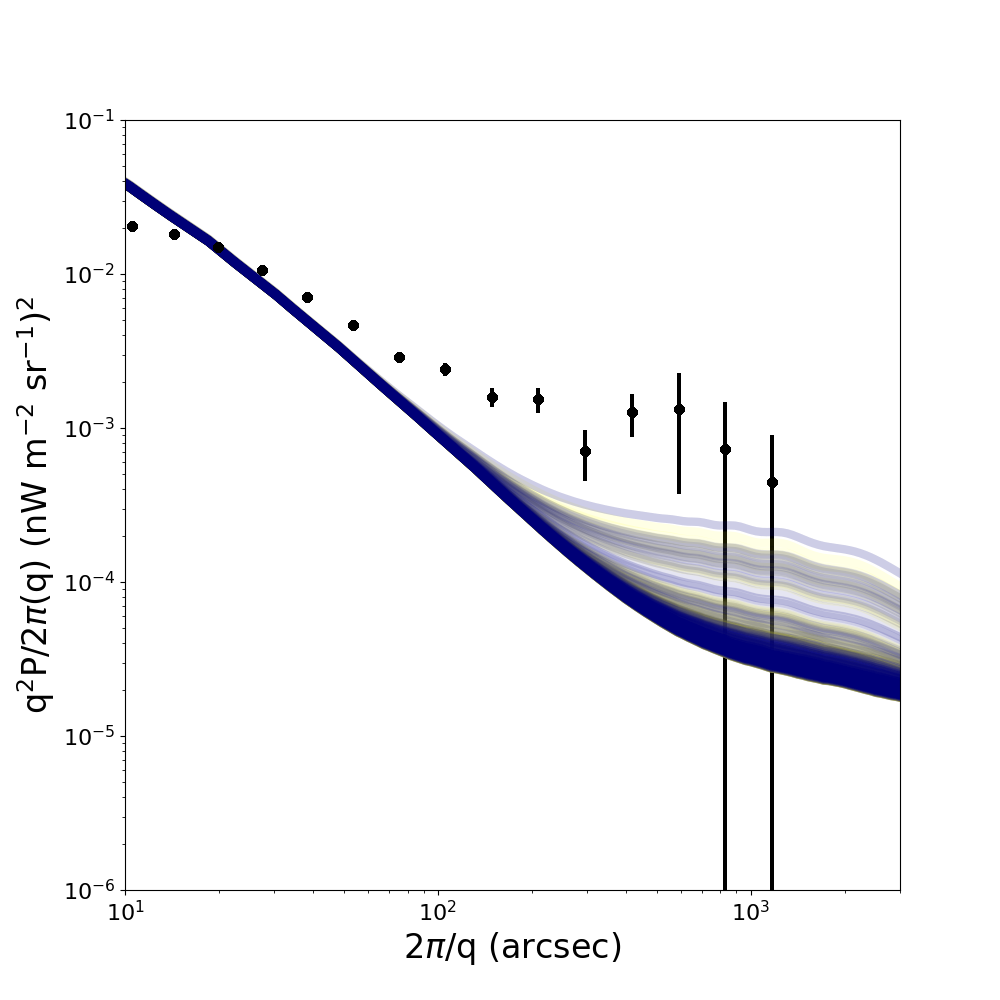}
\includegraphics[width=0.45\textwidth]{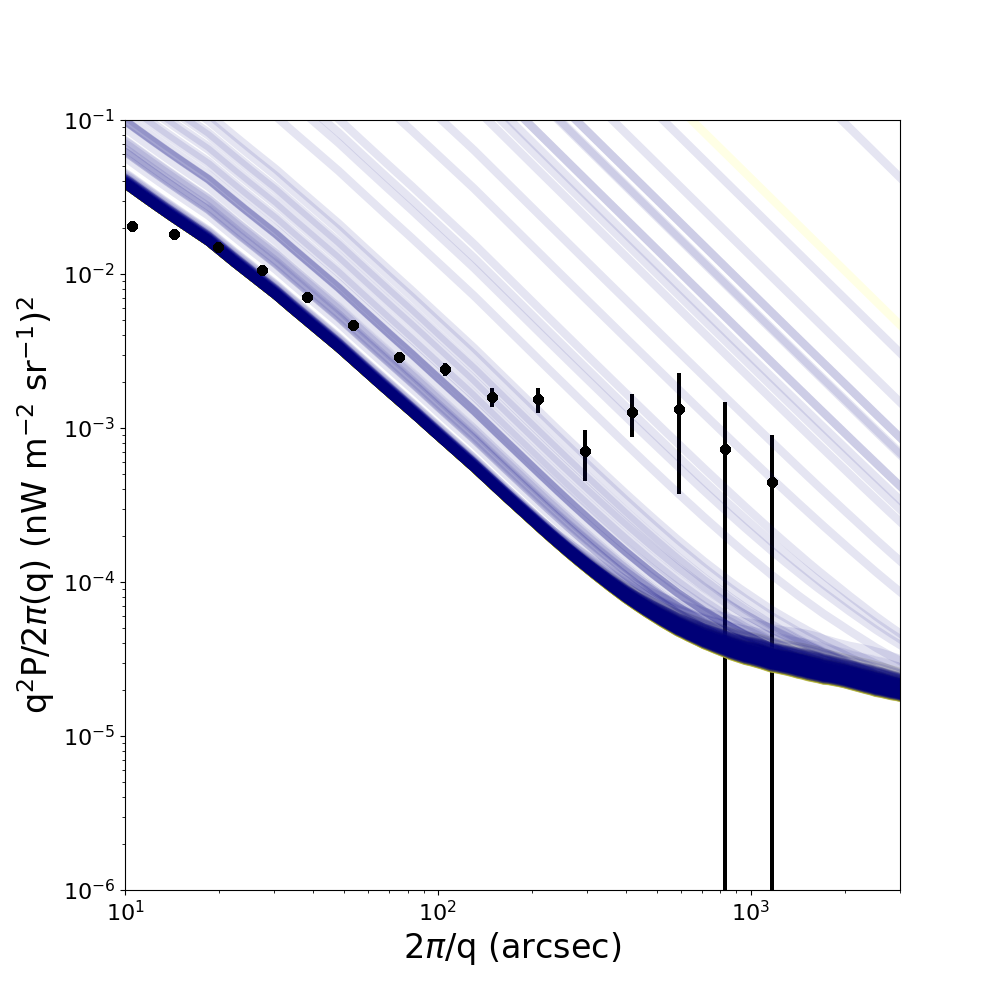}
\caption{he power spectrum of the unresolved [2-5]$\mu$m CIB fluctuations from our model added to the foreground galaxy estimate from \citet{helgason12}. Data points are combination of the results of \citet{2012ApJ...753...63K,li18}. The grey, yellow and navy lines represent {\bf Early, Medium and Late} enrichment, respectively. The two panels represent from $left$ to $right$ the IMF-3B and IMF-3C models. The density of the lines represent the probability of realization of our model based on the sampling of the parameter space. Neither of these two models can fully account for the observed signal.}

\end{figure}
\begin{figure}[h]
\center

\includegraphics[width=0.45\textwidth]{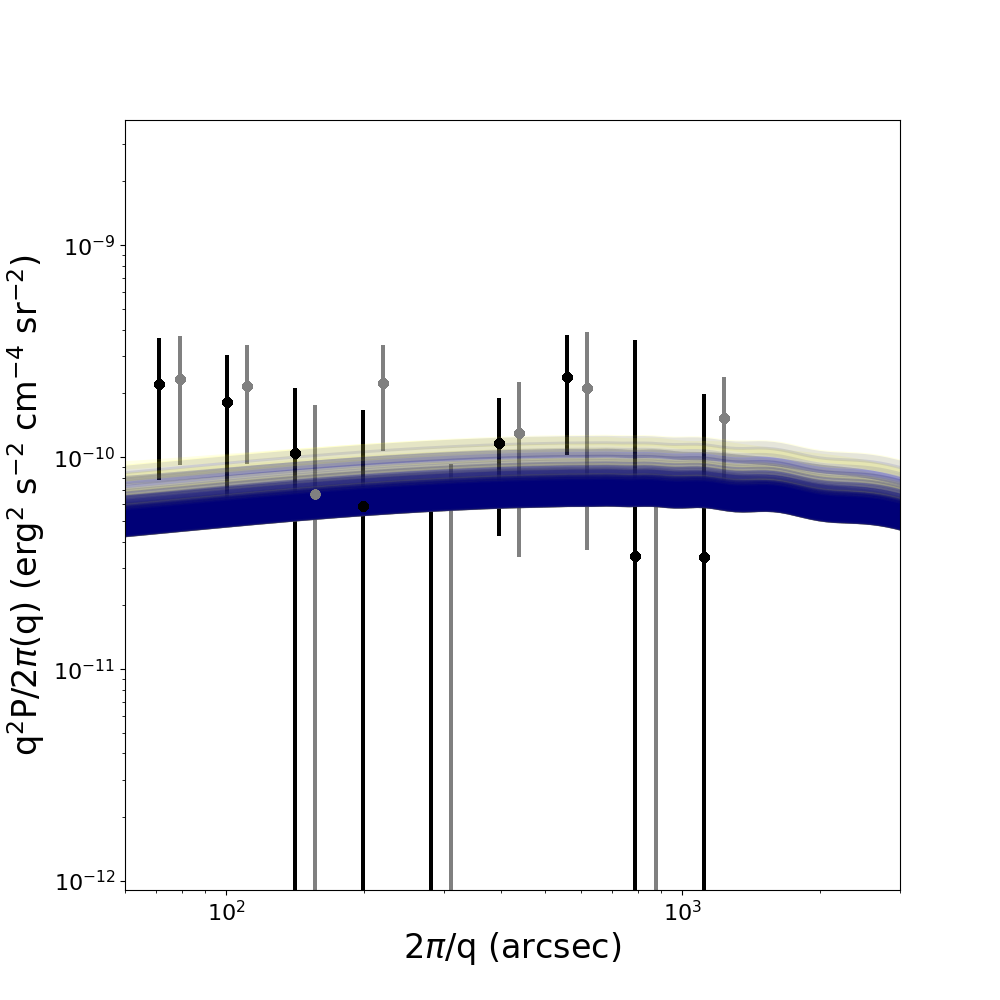}
\includegraphics[width=0.45\textwidth]{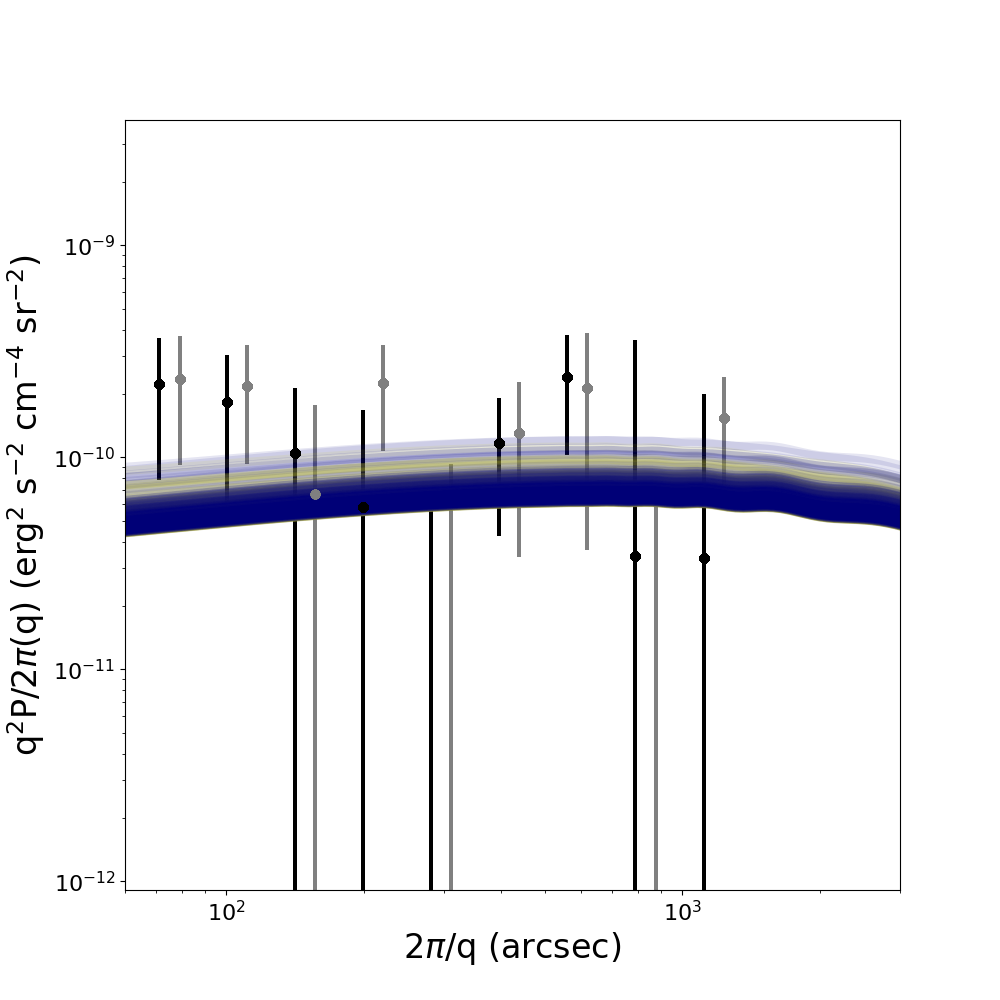}
\caption{The power spectrum of [0.5-2] keV sources modeled here - PBHs + X-ray Binaries -  compared with the expected power from the "unknown" population producing the CIB and CXB joint fluctuations as derived from their coherence by \citet{k19}. The two panels represent from $left$ to $right$ the IMF-3B and IMF-3C models. The grey, yellow and navy lines represent {\bf Early, Medium \& Late} enrichment, respectively. The density of the lines represent the probability of realization of our model based on the sampling of the parameter space.}

\end{figure}

\begin{figure}[h]
\center

\includegraphics[width=0.45\textwidth]{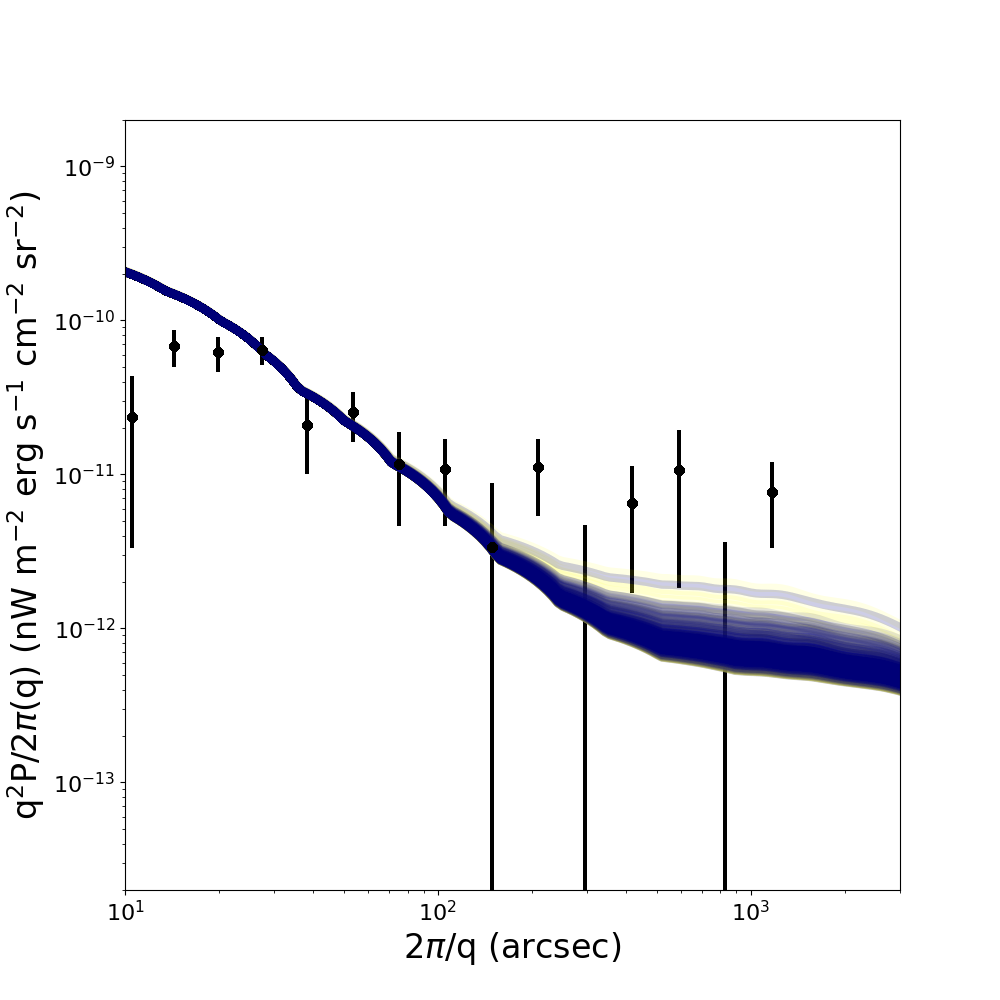}
\includegraphics[width=0.45\textwidth]{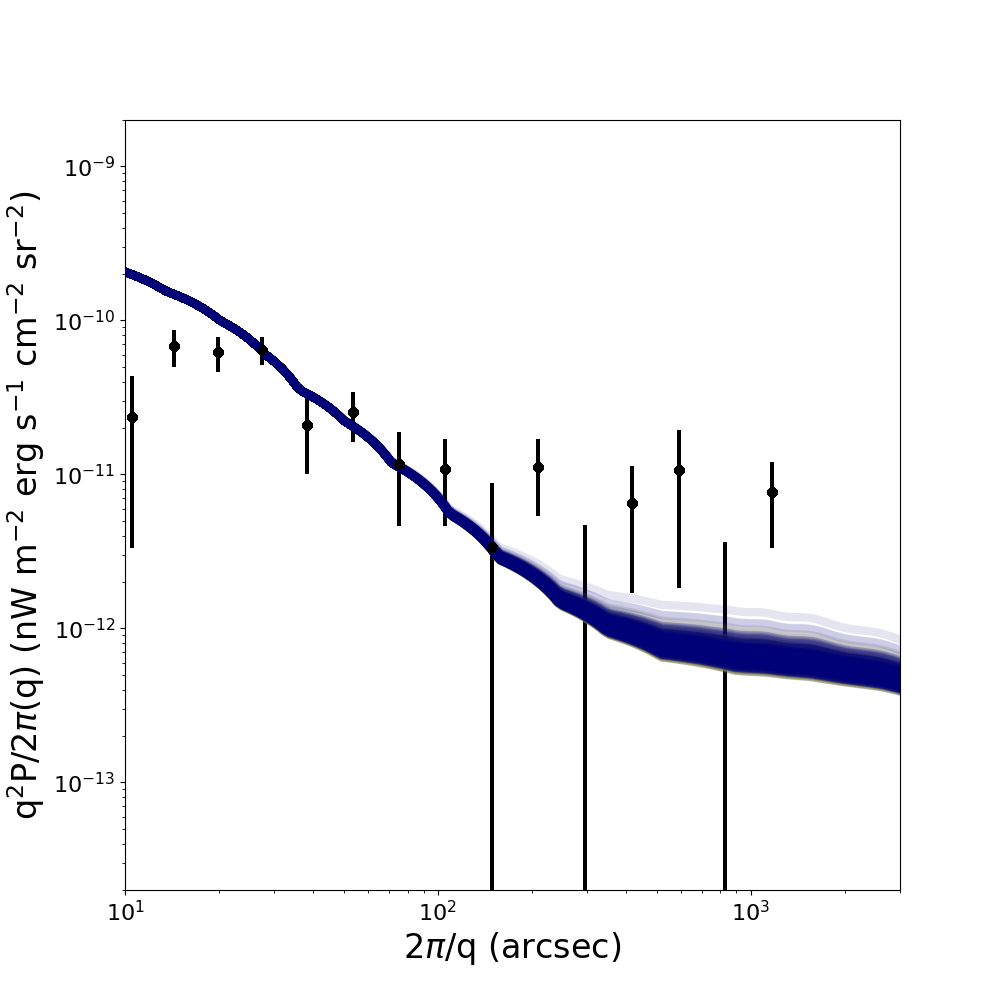}
\caption{The cross-power spectrum of the unresolved [2-5] $\mu$m  CIB and [0.5-2] keV CXB fluctuations from our model added to the foreground galaxy estimate from \citet{helgason12}. Data points are a combination of the results presented in  \citet{cap13,cap17}. The grey, yellow and navy lines represent {\bf Early, Medium \& Late} enrichment, respectively. The two panels represent from $left$ to $right$ the IMF-3B and IMF-3C models. The density of the lines represents the probability of realization of our model based on the sampling of the parameter space. IMF-3C produces too many faint sources hence overproducing the shot-noise.}

\end{figure}

\end{document}